\documentclass[aps,prd,twocolumn,superscriptaddress,nofootinbib,floatfix]{revtex4-2}

\usepackage{graphicx}
\usepackage{amsmath,amssymb,bm}
\usepackage{xcolor}
\colorlet{BLUE}{blue}
\usepackage[colorlinks=true,linkcolor=blue,citecolor=blue,urlcolor=blue]{hyperref}

\graphicspath{{./}}

\DeclareRobustCommand{\rev}[1]{\textcolor{black}{#1}}
\DeclareRobustCommand{\redrev}[1]{\textcolor{black}{#1}}

\usepackage{epstopdf}
\begin{document}
\raggedbottom

\title{Holographic Einstein rings of a covariant quantum corrected AdS black hole}

\author{Zhenglong Ban}
\email{zlban123@163.com}
\affiliation{College of Physics Science and Technology, Hebei University,
  Baoding 071002, China}

\author{Xiaomei Liu}
\email{18866008682@163.com}
\affiliation{College of Physics Science and Technology, Hebei University,
	Baoding 071002, China}
	
\author{Rongjia Yang}
\thanks{Corresponding author}
\email{yangrongjia@tsinghua.org.cn}
\affiliation{College of Physics Science and Technology, Hebei University,
  Baoding 071002, China}
\affiliation{Hebei Key Lab of Optic-Electronic Information and Materials,
  Hebei University, Baoding 071002, China}
\affiliation{National-Local Joint Engineering Laboratory of New Energy
  Photoelectric Devices, Hebei University, Baoding 071002, China}
\affiliation{Key Laboratory of High-precision Computation and Application of
  Quantum Field Theory of Hebei Province, Hebei University,
  Baoding 071002, China}

\begin{abstract}
\rev{\redrev{We investigate holographic Einstein rings} in a covariant quantum corrected
Schwarzschild--AdS$_4$ spacetime using a massless scalar probe. At fixed
horizon radius $r_h$, the deformation parameter $\zeta$ leaves the metric
mass parameter, coordinate surface gravity, and photon sphere radius
unchanged, while increasing the photon potential maximum and the
asymptotic lapse coefficient $A=1+\zeta^2$. The critical coordinate impact parameter approaches $A^{-1/2}$ for large
black holes, whereas the corresponding incidence angle in the normalized
boundary frame approaches $\pi/2$. Ray tracing and local WKB barrier data recover the same
geometric scale. Quasinormal spectra computed with AdS boundary conditions
\redrev{provide a comparison at finite multipole number}. These
results separate the finite radius effect of the deformation from the
change in boundary time normalization.}
\end{abstract}

\maketitle

\section{Introduction}
\label{sec:intro}

The AdS/CFT correspondence \cite{Maldacena:1997re,Gubser:1998bc,Witten:1998qj}
identifies certain strongly coupled quantum field theories with gravitational
dynamics in asymptotically anti--de Sitter (AdS) spacetime. A longstanding
\redrev{question is how to determine from field theory
observables whether a given quantum system admits a gravitational dual and, if
so, what geometry it encodes} \cite{Heemskerk:2009pn}.
Motivated by the first images of supermassive black holes taken by the Event
Horizon Telescope \cite{EHT:2019dse,Falcke:1999pj}, Hashimoto, Kinoshita and
Murata \rev{proposed an imaging construction in which an Einstein ring of the}
dual black hole can be reconstructed from the response function of a boundary
operator subject to a localized source that is periodic in time
\cite{Hashimoto:2019jmw,Hashimoto:2020mgx}.
In this construction, a scalar wave injected at a point of the boundary sphere
propagates through the bulk, is lensed by the black hole, and is read out at
another boundary point; a virtual converging lens applied to the response
function, mathematically a windowed Fourier transform, then produces an
image on a screen, in which the photon sphere of the bulk black hole appears
as a bright ring \cite{Nambu:2012wa,Kanai:2013rga,Nambu:2015aea}.
The idea has since been extended to a variety of settings, including the
optical appearance of the dual of a holographic superconductor probed by
electric currents \cite{Hartnoll:2008vx,Kaku:2021xqp}, charged AdS black holes
\cite{Liu:2022cev}, Gauss--Bonnet AdS black holes \cite{Zeng:2023fqy}, and AdS geometries \rev{motivated by quantum gravity}
  \cite{Wang:2025image,Zeng:2025lqg,Qi:2025as}\rev{.}
\rev{These studies relate the ring morphology and radius to the bulk
geometry and its deformation parameters.}

In parallel, the resolution of black hole singularities
\cite{Penrose:1964wq} remains a central target of quantum gravity. Loop
quantum gravity (LQG) \cite{Thiemann:2007pyv,Rovelli:2004tv} and its
models obtained by symmetry reduction have produced a variety of nonsingular effective
black hole spacetimes \cite{Ashtekar:2018lag}. Many such models, however,
\redrev{break general covariance at the effective level}. \rev{Zhang, Lewandowski, Ma, and Yang constructed effective
Hamiltonian constraints that preserve spacetime covariance
\cite{Zhang:2024khj,Zhang:2024ney}. Zhang and Cao related covariance
conditions to a generalized Birkhoff theorem \cite{Zhang:2025hlm}.
Yang, Zhang, and Ma extended the construction to the electrovacuum
sector with a cosmological constant \cite{Yang:2025ufs}.} The phenomenology
of these covariant quantum corrected black holes, including quasinormal
spectra, shadows, lensing, and thermodynamics, has attracted considerable attention
\cite{Konoplya:2025mvj,Liu:2024soc,Lin:2025fcz,Wang:2025ihs,Liu:2025qnm,
Dong:2026qnm}.

\rev{Our previous study of rotating black holes in this covariant effective
family examined shadow sizes, cusps near extremality, and constraints from
Event Horizon Telescope observations in the absence of a cosmological
constant \cite{Ban:2025epjc}. Here we consider the nonrotating AdS geometry
and its holographic images. The varying asymptotic lapse requires an
explicit distinction between coordinate screen radii and angles measured
in a common boundary frame.}

\rev{For the covariant effective family of
Refs.~\cite{Zhang:2024khj,Zhang:2024ney,Yang:2025ufs}, the asymptotic lapse
varies with the quantum parameter. Comparisons across this family must
therefore specify the boundary time normalization. We consider ``Solution~1'' of}
Ref.~\cite{Yang:2025ufs} in the form presented in Ref.~\cite{Dong:2026qnm}.
After setting the charge to zero and continuing to a negative cosmological
constant, the solution describes a quantum corrected Schwarzschild--AdS$_4$
black hole \redrev{with a single quantum parameter} $\zeta$. This background admits
several useful exact relations:
(i) the event horizon and the surface gravity associated with the coordinate
Killing field $\partial_t$ are independent of $\zeta$;
(ii) the photon sphere radius is also independent of $\zeta$; \rev{and}
(iii) the height of the photon potential depends on $\zeta$ in closed form.
The asymptotic lapse coefficient also changes, so a comparison at fixed
coordinate temperature is not a comparison at fixed temperature in a common,
normalized boundary metric.  We therefore keep the raw screen coordinate and
the incidence angle in the normalized boundary frame distinct throughout the analysis.

Concretely, we place a monochromatic Gaussian source of frequency $\omega$
at the south pole of the boundary two-sphere, compute the response
$\langle O(\theta)\rangle$ of the dual scalar operator by solving the bulk
wave equation, and convert the response into an image with the virtual lens
of Refs.~\cite{Hashimoto:2019jmw,Hashimoto:2020mgx}. \redrev{In geometric optics,}
\rev{the critical coordinate impact parameter is}
\begin{equation}
\rev{b_t\equiv\left.\frac{L}{E_t}\right|_{\rm critical}}
=\big[\,w_{\max}\,(1+\zeta^{2}w_{\max})\,\big]^{-1/2},
\label{eq:ringintro}
\end{equation}
with $w_{\max}$ the peak of the classical Schwarzschild--AdS photon
potential.  In the normalized boundary time $\tau=\sqrt{1+\zeta^2}\,t$,
the invariant angular quantity is
$\sin\vartheta_c=\sqrt{1+\zeta^2}\,b_t$.  The full wave optics computation
tests Eq.\eqref{eq:ringintro} in the coordinate convention used by the lens
and exhibits the resolution limit imposed by the finite aperture.

The paper is organized as follows. Section~\ref{sec:metric} introduces the
background and derives the properties that are independent of $\zeta$.
Section~\ref{sec:scalar} sets up the scalar probe, the boundary expansion in
the presence of the renormalized AdS radius, and the numerical method.
Section~\ref{sec:optics} recalls the imaging formula.
Section~\ref{sec:geodesics} analyzes null geodesics and derives
Eq.\eqref{eq:ringintro}. Section~\ref{sec:images} presents the holographic
images, and Sec.~\ref{sec:radius} the quantitative comparison between the
wave optics screen radius and the geodesic prediction.
Sections~\ref{sec:wkb}--\ref{sec:pseudospectral} develop the WKB,
time domain Prony, backward ray tracing, and pseudospectral calculations;
Sec.~\ref{sec:synthesis} compares their domains of validity and normalization
conventions.
Section~\ref{sec:conclusion} concludes. Numerical details and convergence
tests are collected in the Appendix. \rev{We set the AdS radius of the
classical seed metric to $R=1$ and use $G=\hbar=c=1$.}

\section{Covariant quantum corrected Schwarzschild--AdS$_4$ black hole}
\label{sec:metric}

The covariant effective dynamics of spherically symmetric electrovacuum
gravity with a cosmological constant admits the quantum corrected solution
labeled ``Solution~1'' in Refs.~\cite{Yang:2025ufs,Dong:2026qnm}:
\begin{equation}
\mathrm{d}s^2=-f(r)\,\mathrm{d}t^2+\frac{\mathrm{d}r^2}{f(r)}+r^2\big(\mathrm{d}\theta^2+\sin^2\!\theta\,
\mathrm{d}\varphi^2\big),
\label{eq:metric}
\end{equation}
with
\begin{equation}
f(r)=F_0(r)\left[1+\frac{\zeta^2}{r^2}\,F_0(r)\right],
\label{eq:f}
\end{equation}
where $\zeta$ is the quantum parameter. Setting $Q=0$ and choosing the
branch with a negative cosmological constant $\Lambda=-3/R^2$ with $R=1$,
\begin{equation}
F_0(r)=1-\frac{2M}{r}+r^{2}
\label{eq:F0}
\end{equation}
is the classical Schwarzschild--AdS$_4$ lapse. For $\zeta\to0$ the geometry
reduces to Schwarzschild--AdS$_4$, the background of the original imaging
proposal \cite{Hashimoto:2019jmw,Hashimoto:2020mgx}. \rev{Quasinormal spectra of the de Sitter branch were studied in
Refs.~\cite{Liu:2025qnm,Dong:2026qnm}. The AdS branch has been examined
through thermodynamics \cite{Wang:2025ihs} and tidal Love numbers
\cite{Liu:2025qnm}.} \rev{Microscopically, $\zeta$
originates in the holonomy corrections of effective LQG. The estimate
$\zeta/M\lesssim7.09$ in Refs.~\cite{Lin:2025fcz,Dong:2026qnm} depends on
the underlying mass normalization and minimum mass assumptions. It is not
an independently established bound for the AdS family considered here.
Our principal tables use $r_h\geq0.3$ and $0\leq\zeta\leq1$, for which
$\zeta/M\leq6.12$. Wider scans are phenomenological explorations of the
effective metric.} Since a microscopic boundary dual of the
effective geometry is not known, we adopt the working assumption standard in
holographic studies of modified black holes
\cite{Liu:2022cev,Zeng:2023fqy}: Eq.\eqref{eq:metric} is treated as an
effective bulk background probed by a test scalar, and the standard
holographic dictionary is applied to its asymptotic data.

\rev{The following properties of Eqs.\eqref{eq:f}--\eqref{eq:F0}
determine the photon ring scale and its boundary normalization.}

(i) Horizon and coordinate surface gravity are independent of $\zeta$.
Outside the horizon $F_0>0$, so the bracket in Eq.\eqref{eq:f} obeys
\rev{$1+\zeta^2F_0/r^2\geq1$} and the outermost root of $f$ coincides with that of
$F_0$,
\begin{equation}
M=\tfrac{1}{2}\,r_h\big(1+r_h^{2}\big),
\label{eq:mass}
\end{equation}
for every $\zeta$. An additional outer horizon may appear beyond a
critical $\zeta_c$ in the de Sitter branch \cite{Dong:2026qnm}. No analogous
bound occurs in the AdS branch, for which $f>0$ throughout the exterior
$r>r_h$.
Moreover, because the bracket equals unity at $F_0=0$,
\begin{equation}
f'(r_h)=F_0'(r_h)\quad\Longrightarrow\quad
T=\frac{f'(r_h)}{4\pi}=\frac{3r_h^{2}+1}{4\pi r_h},
\label{eq:temperature}
\end{equation}
\redrev{which equals the Schwarzschild--AdS$_4$ temperature when the Killing field is normalized as}
$\partial_t$.  This is a coordinate Hawking temperature, denoted $T_t$
below.  Because the normalization of $\partial_t$ changes at infinity when
$\zeta$ varies, the temperature in a common boundary frame is not invariant;
Eq.\eqref{eq:Tbdy} gives the normalized result.\footnote{\rev{For the
classical Schwarzschild--AdS seed, the heat capacity changes sign at
$r_h=R/\sqrt{3}$, whereas the Hawking--Page transition occurs at $r_h=R$
\cite{Hawking:1982dh}. These are distinct thresholds. The present probe
calculation does not determine the thermodynamic potential of the
quantum corrected effective theory.}}

(ii) The effective AdS radius is renormalized.
Expanding Eq.\eqref{eq:f} at large $r$,
\begin{equation}
f=A\,r^{2}+B-\frac{2MB}{r}+\frac{\zeta^{2}}{r^{2}}
-\frac{4M\zeta^{2}}{r^{3}}+\mathcal{O}(r^{-4}),
\label{eq:asym}
\end{equation}
with
\begin{equation}
A\equiv1+\zeta^{2},\qquad B\equiv1+2\zeta^{2}.
\label{eq:AB}
\end{equation}
The geometry is asymptotically AdS$_4$ with effective cosmological constant
$\Lambda_{\rm eff}=-3A$ and effective AdS radius
\begin{equation}
L_{\rm eff}=\big(1+\zeta^{2}\big)^{-1/2}.
\label{eq:Leff}
\end{equation}
The conformal boundary metric is
$\mathrm{d}s^2_{\rm bdy}\propto -A\,\mathrm{d}t^{2}+\mathrm{d}\Omega_2^{2}$, so that the proper boundary
time in the unit sphere conformal frame is $\tau=\sqrt{A}\,t$.  Frequencies
and temperatures in that frame are
\begin{equation}
\omega_{\rm bdy}=\frac{\omega_t}{\sqrt{A}},\qquad
T_{\rm bdy}=\frac{T_t}{\sqrt{A}}.
\label{eq:Tbdy}
\end{equation}
The imaging calculation is performed in $t$ with frequency $\omega_t$,
as in the original holographic lens convention. Its output is therefore a
coordinate screen scale, which is converted explicitly when a normalized
boundary angle is quoted. The factor $\sqrt{A}$ remains in this conversion.

(iii) The exterior is a monotone deformation of Schwarzschild--AdS.
Let $w(r)\equiv F_0(r)/r^{2}$ denote the classical photon potential
introduced in Sec.~\ref{sec:geodesics}. The full potential is the pointwise
monotone map $v=w+\zeta^{2}w^{2}$, which leads to the closed form ring radius
derived below.

\section{Scalar probe and boundary response}
\label{sec:scalar}

Following Refs.~\cite{Hashimoto:2019jmw,Hashimoto:2020mgx}, we probe the
geometry with a massless scalar $\Phi$, dual to a boundary operator $O$ of
conformal dimension $\Delta_O=3$, sourced by an axisymmetric, monochromatic
Gaussian localized at the south pole of the boundary $S^{2}$,
\begin{align}
J_O(t,\theta)&=e^{-i\omega t}\,\rev{j(\theta)},\nonumber\\
\rev{j(\theta)}&=\frac{1}{2\pi\sigma^{2}}\,
\exp\!\left[-\frac{(\pi-\theta)^{2}}{2\sigma^{2}}\right],
\label{eq:source}
\end{align}
with $\sigma\ll\pi$. In ingoing Eddington--Finkelstein coordinates
$v=t+r_*$, $\mathrm{d}r_*=\mathrm{d}r/f$, and with $u\equiv1/r$, the Klein--Gordon equation
$\Box\Phi=0$ separates as
$\Phi=e^{-i\omega v}\sum_{l}c_l\,U_l(u)\,Y_{l0}(\theta)$ with
\begin{align}
&\rev{u^{2}g\,U_l''+\big(u^{2}g'-2ug+2i\omega u^{2}\big)U_l'}
\nonumber\\
&\qquad\qquad\rev{-\big(2i\omega u+l(l+1)\,u^{2}\big)U_l=0,}
\label{eq:radial}
\end{align}
\rev{where $g(u)\equiv u^2f(1/u)$ and primes denote $u$ derivatives.
The signs of the frequency terms follow from $v=t+r_*$ and
$e^{-i\omega v}$.} Regularity at the horizon and a unit Dirichlet source
$U_l(0)=1$ at the boundary fix the solution. The expansion coefficients of
the source \eqref{eq:source} in spherical harmonics are
$c_l\simeq(-1)^{l}\sqrt{(l+\tfrac12)/2\pi}\,
e^{-(l+1/2)^{2}\sigma^{2}/2}$ \cite{Hashimoto:2020mgx}.

\rev{The boundary expansion must account for the asymptotic coefficient
$A$ in Eq.\eqref{eq:asym}. Writing
$g(u)=g_0+g_2u^2+g_3u^3+\cdots$, one obtains}
\begin{equation}
\rev{g_0=A,\qquad g_1=0,\qquad g_2=B,\qquad g_3=-2MB.}
\label{eq:fu}
\end{equation}
Substituting a series $U_l=1+a_1u+a_2u^{2}+a_3u^{3}+\cdots$ into
Eq.\eqref{eq:radial} gives, order by order,
\begin{equation}
\rev{a_1=-\frac{i\omega}{g_0},\qquad
a_2=-\frac{l(l+1)}{2g_0},}
\label{eq:a1a2}
\end{equation}
\rev{while the equation at order $u^3$ imposes no condition on $a_3$:
$2i\omega a_2-l(l+1)a_1=0$. Thus}
\begin{equation}
\rev{U_l=1-\frac{i\omega}{g_0}\,u-\frac{l(l+1)}{2g_0}\,u^{2}
+\langle O\rangle_l\,u^{3}+\mathcal{O}(u^{4})}
\label{eq:bexp}
\end{equation}
\rev{for $g_1=0$, with $g_0\ne0$ and arbitrary $g_2,g_3$. The coefficient
$a_3\equiv\langle O\rangle_l$ is the normalizable response datum in this
coordinate convention. Its conversion to the renormalized boundary
one-point function can add local terms built from the source and its
derivatives. For the narrow south pole source, these terms are negligible
in the observation patches used here, which are separated from the source.
We therefore assemble the image from $a_3$, following
Refs.~\cite{Liu:2022cev,Zeng:2023fqy}:}
\begin{equation}
\langle O(\theta)\rangle
=\sum_{l=0}^{l_{\max}}c_l\,\langle O\rangle_l\,Y_{l0}(\theta),
\label{eq:response}
\end{equation}
The overall normalization of the response is immaterial because each
image is normalized independently.

Eq.\eqref{eq:radial} is discretized on a Chebyshev--Lobatto grid
of order $N$ in $u\in[0,u_h]$ \cite{Boyd:2000,Jansen:2017oag}. The horizon
and boundary rows impose the regularity and Dirichlet conditions,
respectively. The response is extracted by projecting
\rev{$U_l-[1-i\omega u/g_0-l(l+1)u^{2}/(2g_0)]$ onto $u^{3}$ near the boundary.}
Because the $l$ dependence is confined to the diagonal of the discrete
operator, a single eigendecomposition yields all multipoles; selected modes
are also evaluated with direct LU solves. We use
$N=\lceil\pi\omega/r_h\rceil$ and $l_{\max}=860$ for $\sigma=0.01$.
Convergence tests and Schwarzschild--AdS benchmarks are reported in
Appendix~\ref{app:numerics}.

\section{Imaging with wave optics}
\label{sec:optics}

An observer at $(\theta,\varphi)=(\theta_{\rm obs},0)$ images the response
through a virtual converging lens of radius $d$ and focal length $f_{\rm L}$
\cite{Hashimoto:2019jmw,Hashimoto:2020mgx}. With planar coordinates
$\vec{x}$ on the aperture \rev{defined after rotating the observation
point to the pole}, and $\vec{x}_S$ on a spherical screen, the image amplitude
is the windowed Fourier transform
\begin{equation}
\Psi_S(\vec{x}_S)=\int_{|\vec{x}|<d}\mathrm{d}^{2}x\;
\langle O(\vec{x})\rangle\;
e^{-i\frac{\omega}{f_{\rm L}}\vec{x}\cdot\vec{x}_S},
\label{eq:lens}
\end{equation}
and the image intensity is $|\Psi_S|^{2}$.  Here and below
$\omega\equiv\omega_t$ is conjugate to the bulk coordinate time, so
Eq.\eqref{eq:lens} defines a coordinate screen. A plane wave component
$\propto e^{il\theta}$ of the response produces a peak at
$x_S/f_{\rm L}\simeq l/\omega$.  A lens calibrated to normalized boundary
time instead uses $\omega_{\rm bdy}=\omega/\sqrt{A}$ and assigns the same
multipole the normalized coordinate
$\hat x_S/f_{\rm L}=\sqrt{A}\,x_S/f_{\rm L}$.
The window of radius $d$ limits the coordinate screen resolution to
\begin{equation}
\Delta\!\left(\frac{x_S}{f_{\rm L}}\right)\simeq\frac{\pi}{\omega d},
\label{eq:resolution}
\end{equation}
a scale that will play a physical role in Sec.~\ref{sec:radius}. Unless
stated otherwise we use the fiducial parameters of
Ref.~\cite{Hashimoto:2020mgx}: $\omega=80$, $\sigma=0.01$, $d=0.5$,
$f_{\rm L}=1$, for which $\pi/(\omega d)\simeq0.079$.

\section{Null geodesics and the photon sphere}
\label{sec:geodesics}

\rev{For equatorial null geodesics, let $E_t=f\dot t$ and
$L=r^{2}\dot\varphi$ denote the conserved energy and angular momentum.
The radial equation is $\dot r^{2}=E_t^{2}-L^{2}v(r)$, with}
\begin{align}
v(r)&=\frac{f(r)}{r^{2}}=w(r)+\zeta^{2}w(r)^{2},\nonumber\\
w(r)&\equiv\frac{F_0(r)}{r^{2}}=1+\frac{1}{r^{2}}-\frac{2M}{r^{3}} .
\label{eq:veff}
\end{align}
Because $v$ is a monotonically increasing function of $w$ wherever $w>0$,
\begin{equation}
\frac{\mathrm{d}v}{\mathrm{d}r}=\big(1+2\zeta^{2}w\big)\frac{\mathrm{d}w}{\mathrm{d}r},
\label{eq:chain}
\end{equation}
the extrema of $v$ outside the horizon coincide with those of $w$. Hence the
photon sphere radius is exactly the classical one,
\begin{equation}
r_{\rm ps}=3M=\tfrac{3}{2}\,r_h\big(1+r_h^{2}\big),
\label{eq:rps}
\end{equation}
independent of $\zeta$, \redrev{while the maximum of the potential increases},
\begin{align}
v_{\max}&=w_{\max}\big(1+\zeta^{2}w_{\max}\big),\nonumber\\
w_{\max}&=\frac{(3r_h^{2}+4)(3r_h^{2}+1)^{2}}
{27\,r_h^{2}\,(r_h^{2}+1)^{2}},
\label{eq:vmax}
\end{align}
with $w_{\max}$ the Schwarzschild--AdS value \cite{Hashimoto:2020mgx}.

\rev{The maximum $v_{\max}$ determines the critical null orbit and the
local high multipole barrier frequency,
$\sqrt{l(l+1)v_{\max}}$ \cite{Ferrari:1984zz,Berti:2009kk,Hashimoto:2020mgx}.
Matching this scale to a driving frequency $\omega$ gives
$l_*\simeq\omega/\sqrt{v_{\max}}$. The lens maps that multipole to
$x_S/f_{\rm L}\simeq l_*/\omega$. This is a geometric optics estimate of the
ring position; the global wave intensity maximum can differ at finite
frequency. The critical coordinate impact parameter is}
\begin{equation}
\rev{b_t\equiv\left.\frac{L}{E_t}\right|_{\rm critical}}
=\frac{1}{\sqrt{v_{\max}}}
=\Big[\,w_{\max}\big(1+\zeta^{2}w_{\max}\big)\Big]^{-1/2}.
\label{eq:ring}
\end{equation}
\rev{Since $w_{\max}>1$ and grows as $r_h^{-2}$ for small black holes,
$b_t$ decreases monotonically with $\zeta$. Its fractional decrease is
larger at smaller $r_h$, as follows from}
\begin{equation}
\frac{b_t(\zeta)}{b_t(0)}
=\big(1+\zeta^{2}w_{\max}\big)^{-1/2}.
\label{eq:ratio}
\end{equation}
\rev{The} incidence angle in the unit sphere boundary frame is
\begin{align}
\sin\vartheta_c&=\sqrt{A}\,b_t
=\left[\frac{A}{w_{\max}(1+\zeta^2w_{\max})}\right]^{1/2},\nonumber\\
A&=1+\zeta^2 .
\label{eq:boundaryangle}
\end{align}
Its fractional change is therefore
\begin{equation}
\frac{\sin\vartheta_c(\zeta)}{\sin\vartheta_c(0)}
=\left[\frac{1+\zeta^2}{1+\zeta^2w_{\max}}\right]^{1/2}.
\label{eq:boundaryratio}
\end{equation}
Since $w_{\max}\to1^{+}$ as $r_h\to\infty$, the two conventions have
different limits for large black holes:
\begin{equation}
 b_t\to A^{-1/2}=L_{\rm eff},\qquad
 \sin\vartheta_c\to1 .
\label{eq:saturation}
\end{equation}
\rev{The coordinate screen angle therefore saturates at
$\arcsin L_{\rm eff}$, while the normalized critical ray becomes tangent
to the boundary sphere.} For example, at
$r_h=0.6$, $\zeta=1$, the coordinate lens gives
$x_S/f_{\rm L}=0.600$, consistent with $b_t=0.607$; the same critical ray has
$\sin\vartheta_c=0.858$ after boundary normalization.

The monotonic relation between $v$ and $w$ gives both the extremum
location \eqref{eq:rps} and the closed form \eqref{eq:ring} directly.
Numerical checks of their implementation are summarized in
Appendix~\ref{app:numerics}.

\section{Holographic images}
\label{sec:images}

\begin{figure*}[t]
\centering
\includegraphics[width=0.32\textwidth]{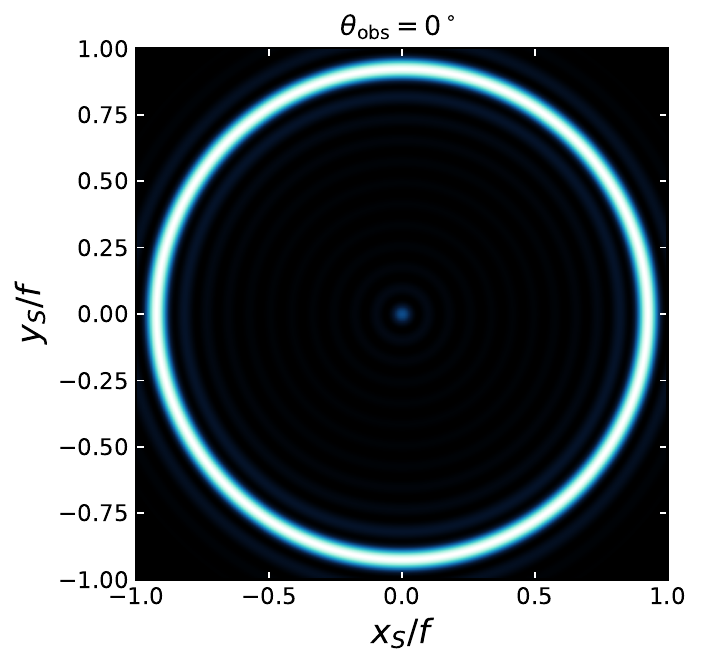}\hfill
\includegraphics[width=0.32\textwidth]{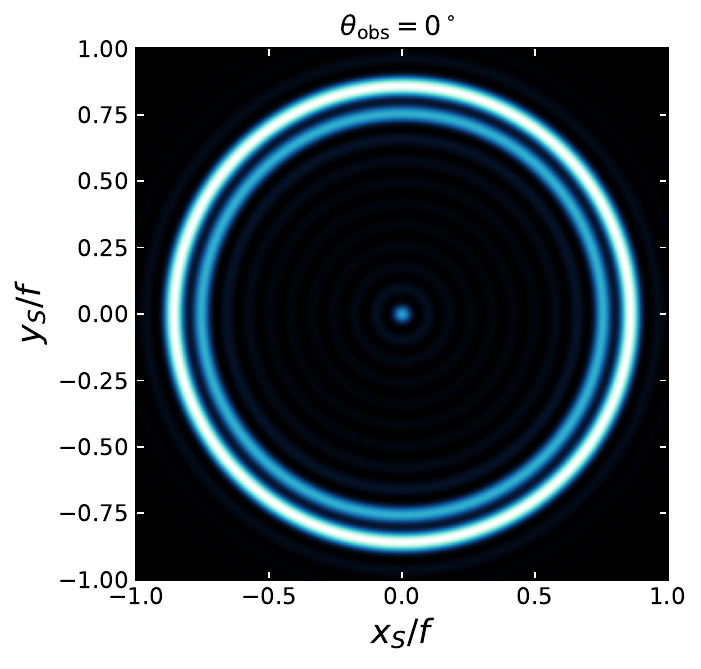}\hfill
\includegraphics[width=0.32\textwidth]{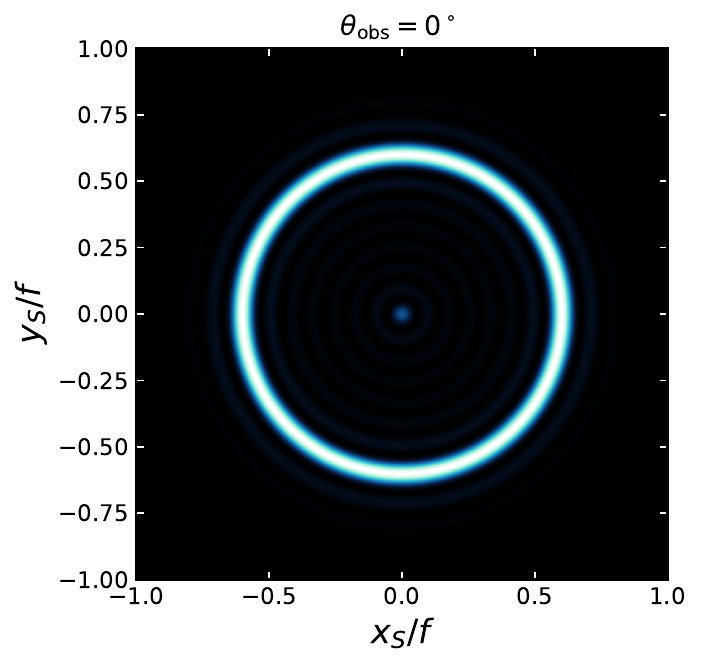}
\caption{Holographic Einstein ring of the covariant quantum corrected
Schwarzschild--AdS$_4$ black hole observed at the north pole
($\theta_{\rm obs}=0^{\circ}$) for $r_h=0.6$, $\omega=80$, $\sigma=0.01$,
$d=0.5$. From left to right: $\zeta=0$, $0.5$, $1$. The horizon radius and
the \rev{coordinate temperature $T_t$} are identical in the three panels; the ring radius
shrinks from $x_S/f_{\rm L}\simeq0.92$ to $0.86$ to $0.60$, following
Eq.\eqref{eq:ring}.}
\label{fig:rings_zeta}
\end{figure*}

\begin{figure*}[t]
\centering
\includegraphics[width=0.24\textwidth]{ring_qSAdS_rh0p6_a1_w80_d0p5_fL1_sig0p01_th0deg.pdf}\hfill
\includegraphics[width=0.24\textwidth]{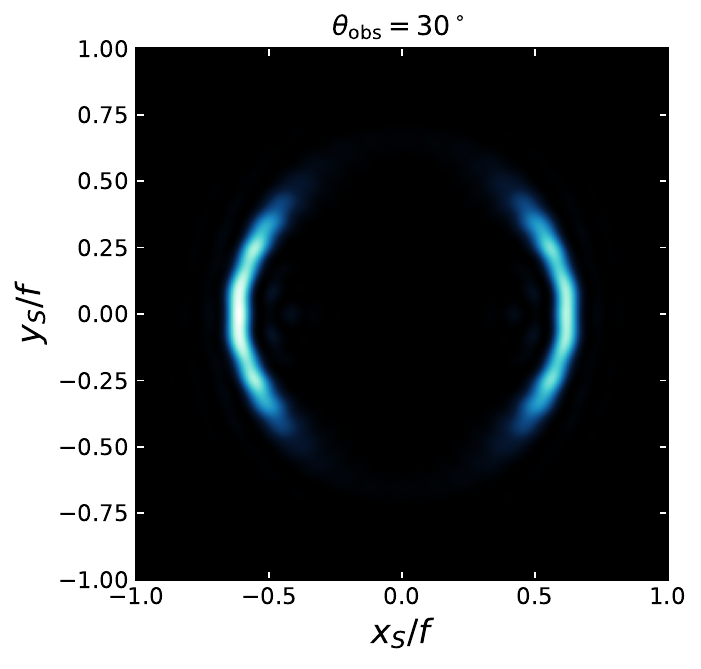}\hfill
\includegraphics[width=0.24\textwidth]{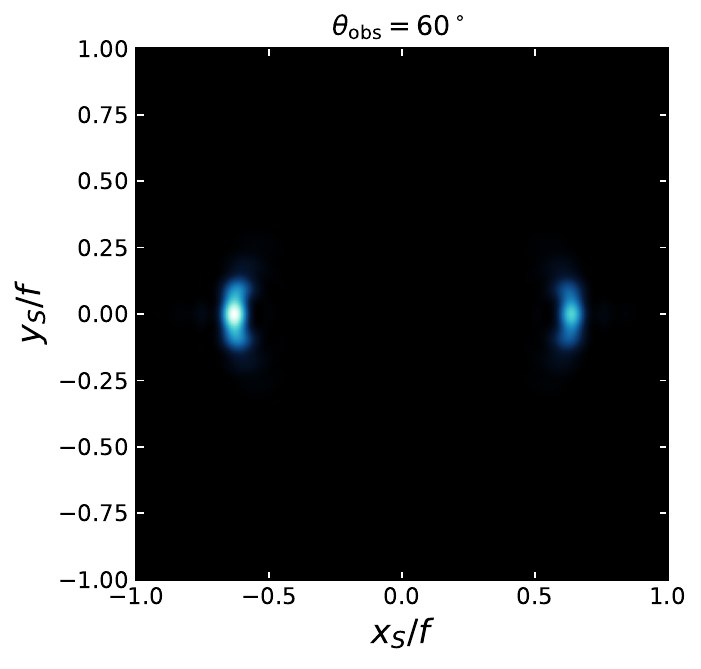}\hfill
\includegraphics[width=0.24\textwidth]{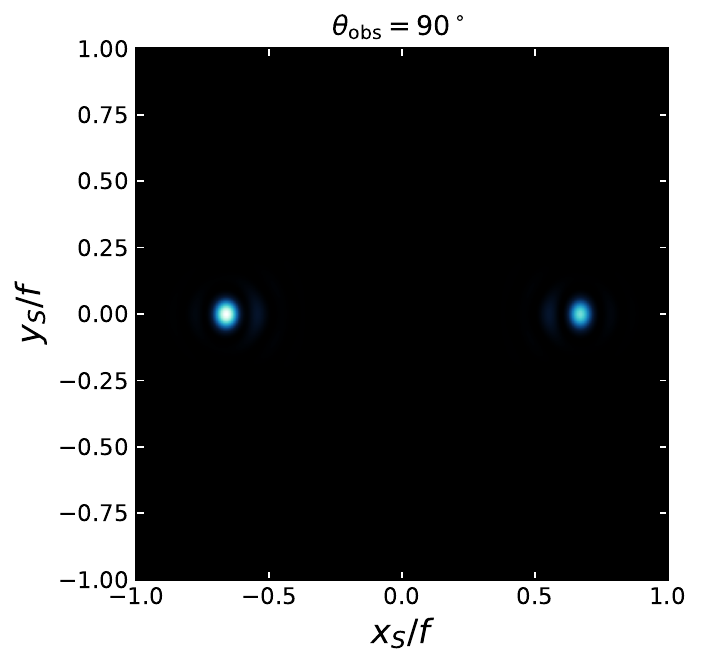}
\caption{Dependence on the observation angle for $\zeta=1$, $r_h=0.6$,
$\omega=80$. From left to right: $\theta_{\rm obs}=0^{\circ}$, $30^{\circ}$,
$60^{\circ}$, $90^{\circ}$. As in the classical case
\cite{Hashimoto:2020mgx}, the ring deforms into arcs and, at
$\theta_{\rm obs}=90^{\circ}$, into a double image of light rays winding
clockwise and counterclockwise around the black hole.
\rev{The characteristic image scale is reduced in the coordinate screen.}}
\label{fig:rings_angle}
\end{figure*}

\begin{figure*}[t]
\centering
\includegraphics[width=0.32\textwidth]{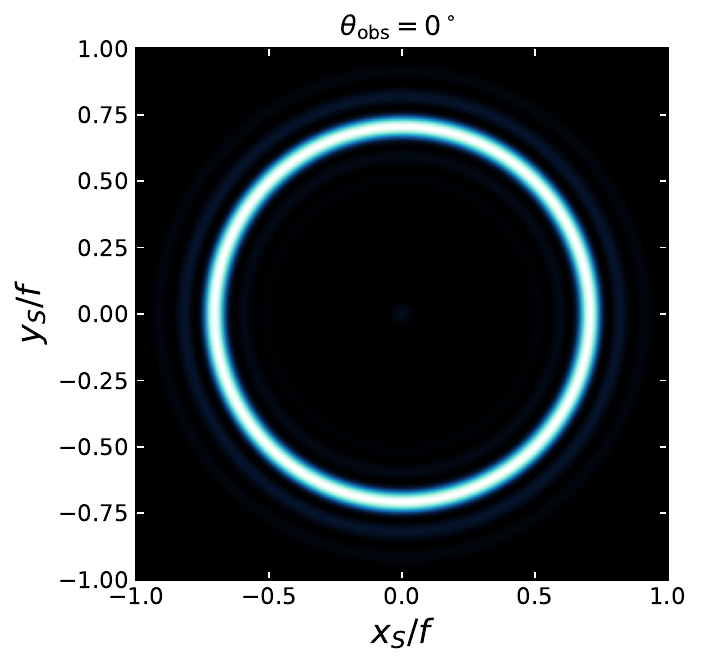}\hfill
\includegraphics[width=0.32\textwidth]{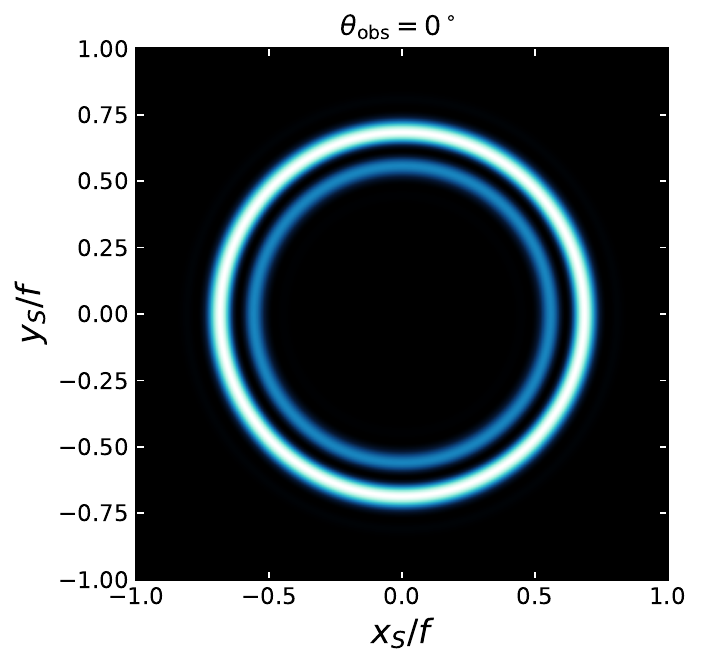}\hfill
\includegraphics[width=0.32\textwidth]{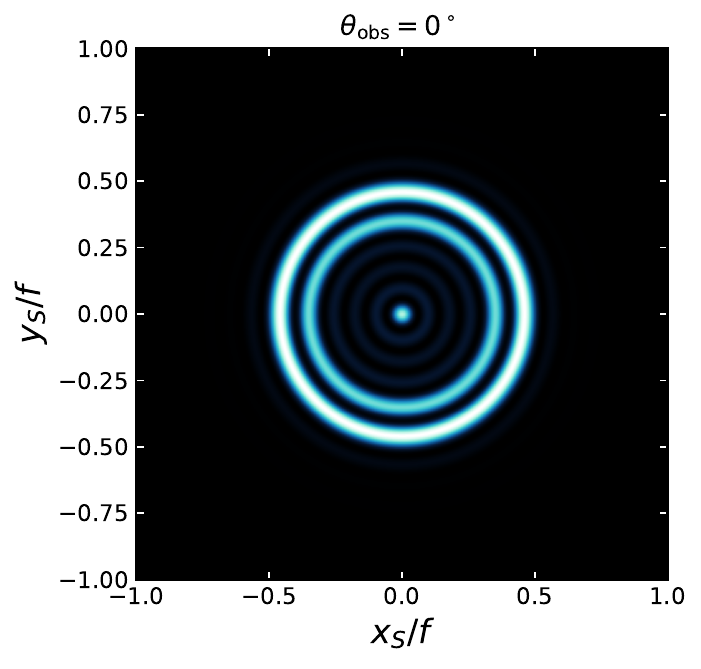}
\caption{Same as Fig.~\ref{fig:rings_zeta} for a smaller black hole,
$r_h=0.3$. From left to right: $\zeta=0$, $0.5$, $1$. The fractional
shrinking of the ring is stronger than at $r_h=0.6$, as predicted by
Eq.\eqref{eq:ratio} with the larger $w_{\max}$.}
\label{fig:rings_rh03}
\end{figure*}

\begin{figure*}[t]
\centering
\includegraphics[width=0.32\textwidth]{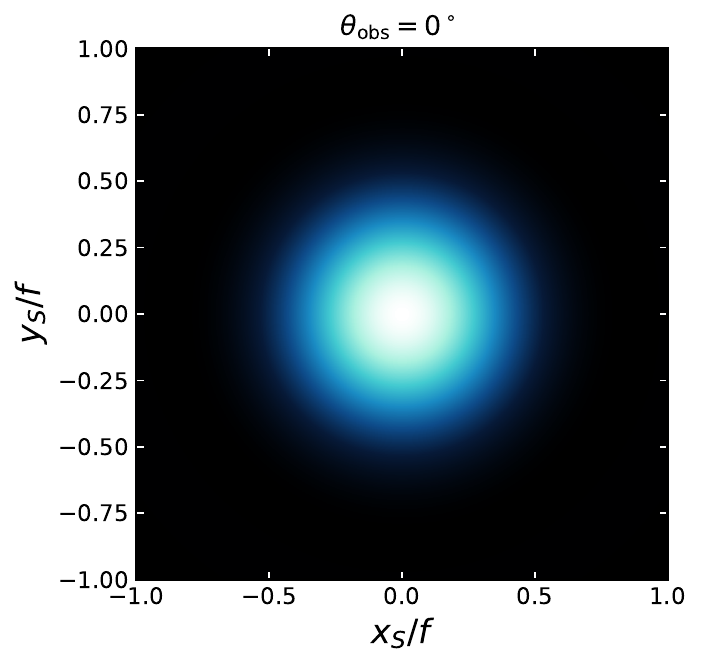}\hfill
\includegraphics[width=0.32\textwidth]{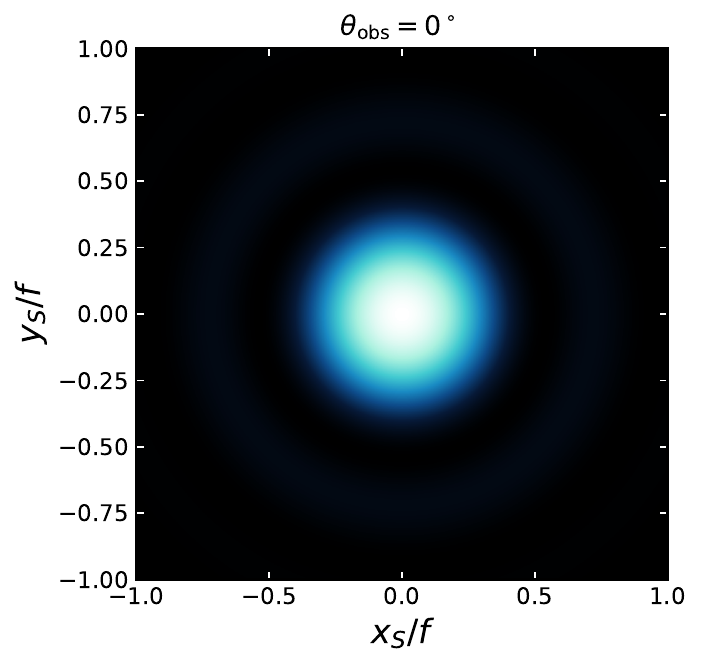}\hfill
\includegraphics[width=0.32\textwidth]{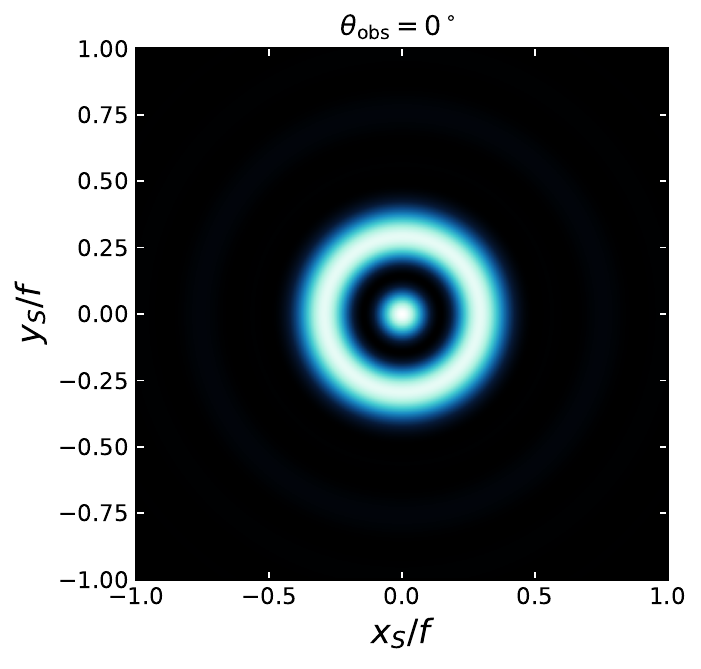}
\caption{Frequency dependence of the image for $\zeta=1$, $r_h=0.3$,
$\theta_{\rm obs}=0^{\circ}$. From left to right: $\omega=10$, $20$, $30$.
As the frequency decreases the geometric optics ring is progressively
blurred and finally washed out by wave effects, in parallel with the
classical behavior \cite{Hashimoto:2020mgx}; note that the resolution scale
\eqref{eq:resolution} grows as $1/\omega$.}
\label{fig:freq}
\end{figure*}

\begin{figure*}[tbp]
\centering
\includegraphics[width=0.48\textwidth]{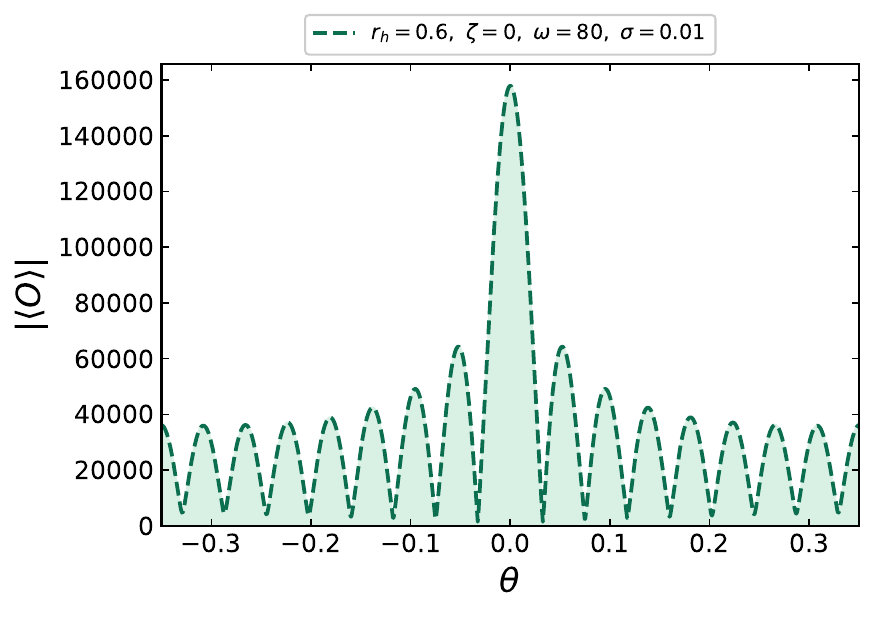}\hfill
\includegraphics[width=0.48\textwidth]{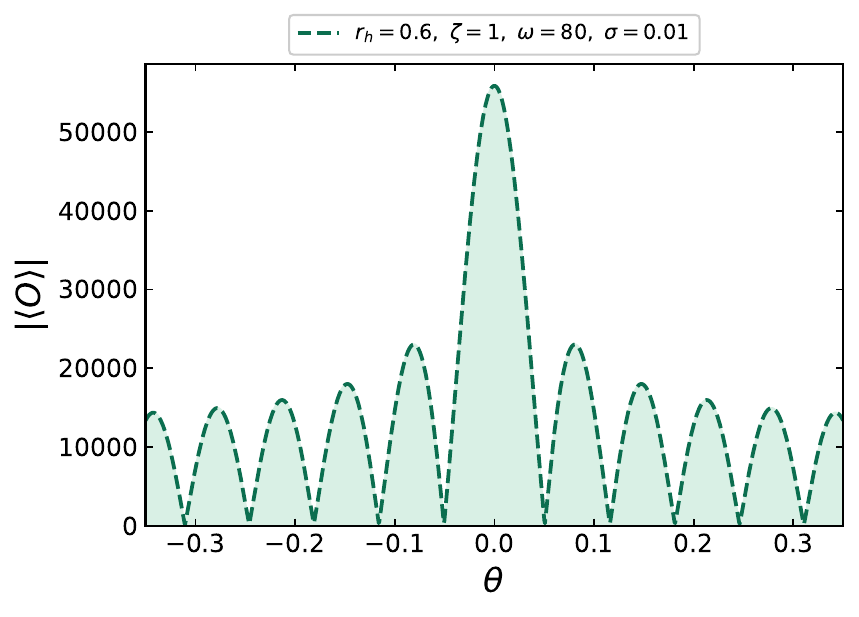}\\[2mm]
\includegraphics[width=0.48\textwidth]{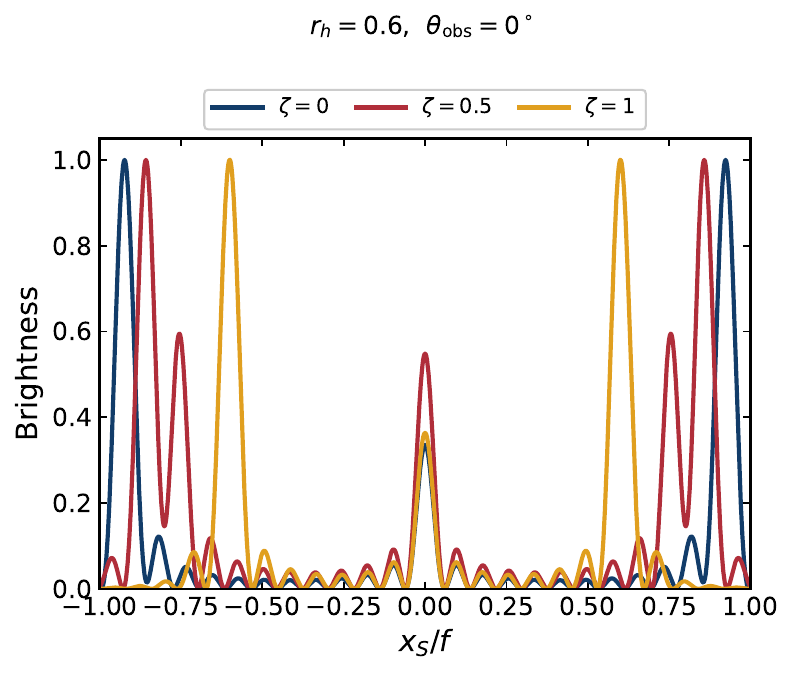}\hfill
\includegraphics[width=0.48\textwidth]{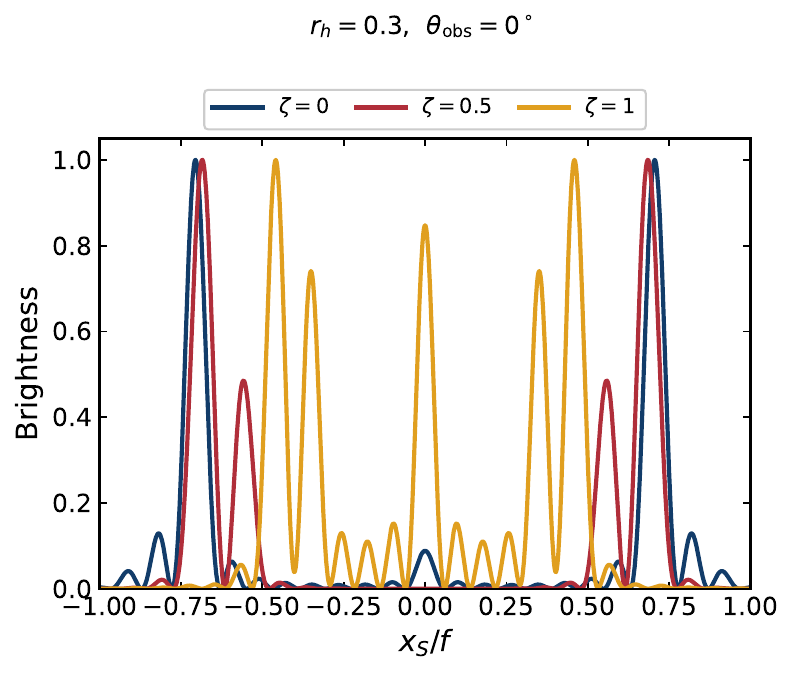}
\caption{Upper row: amplitude of the boundary response
$|\langle O(\theta)\rangle|$ near the image point for $r_h=0.6$,
$\omega=80$, and $\sigma=0.01$. The left and right panels correspond to
$\zeta=0$ and $\zeta=1$, respectively. \rev{The response is defined in
Sec.~\ref{sec:scalar}.} The wider interference fringes
at $\zeta=1$ reflect the smaller dominant multipole
$l_*\simeq\omega/\sqrt{v_{\max}}$. Lower row: normalized screen brightness
at $\theta_{\rm obs}=0^{\circ}$ for $\zeta=0,0.5,1$. The left and right
panels show $r_h=0.6$ and $r_h=0.3$, respectively. \rev{Each brightness
profile is normalized to its own maximum. Prominent noncentral peaks
identify the ring and shift inward as $\zeta$ increases. A central
diffraction maximum can coexist with the ring.}}
\label{fig:response}
\label{fig:brightness}
\end{figure*}

Fig~\ref{fig:rings_zeta} shows the images observed at the north pole,
which is antipodal to the source, for $r_h=0.6$ and
$\zeta=0,\,0.5,\,1$. The three black holes have the same horizon radius and
coordinate temperature $T_t=(3r_h^2+1)/(4\pi r_h)$, whereas the bright ring
contracts from $x_S/f_{\rm L}\simeq0.92$ at $\zeta=0$ to $0.60$ at
$\zeta=1$. Eq.\eqref{eq:ring} gives $0.904$, $0.792$, and $0.607$ for
$\zeta=0$, $0.5$, and $1$, respectively. The remaining offsets arise from
wave optics effects and are quantified in Sec.~\ref{sec:radius}. The
subsidiary concentric fringes are produced by the finite aperture.
\rev{The central intensity also depends on interference within the
aperture and should not be assigned to a unique null trajectory.}

Fig~\ref{fig:rings_angle} displays the $\theta_{\rm obs}$ dependence
for $\zeta=1$. The Schwarzschild--AdS morphology persists: a complete ring
at the pole develops into arcs at intermediate inclination and a double
image at $\theta_{\rm obs}=90^{\circ}$, where the two components correspond
to opposite winding directions. \rev{The characteristic image scale is
smaller in the coordinate screen, although finite frequency morphology
need not undergo a uniform rescaling.} The smaller black hole with $r_h=0.3$, shown in
Fig.~\ref{fig:rings_rh03}, exhibits a larger fractional contraction, as
expected from Eq.\eqref{eq:ratio}. The measured positions
$0.71,\,0.68,\,0.46$ for $\zeta=0,\,0.5,\,1$ exceed the corresponding
geodesic values $0.65,\,0.51,\,0.35$, indicating stronger wave optics
offsets. Fig~\ref{fig:freq} shows the frequency dependence at
$(r_h,\zeta)=(0.3,1)$. As $\omega$ decreases, the eikonal hierarchy weakens
and the aperture resolution \eqref{eq:resolution} becomes coarser, causing
the ring to broaden and eventually disappear.

The boundary response profiles behind these images are shown in
Fig.~\ref{fig:response}, and the brightness slices through the screen center
in Fig.~\ref{fig:brightness}; \rev{these profiles show the inward shift of
the dominant noncentral peak at fixed horizon radius.}

\section{Screen radius: wave optics versus geodesics}
\label{sec:radius}

\begin{figure*}[tbp]
\centering
\includegraphics[width=0.48\textwidth]{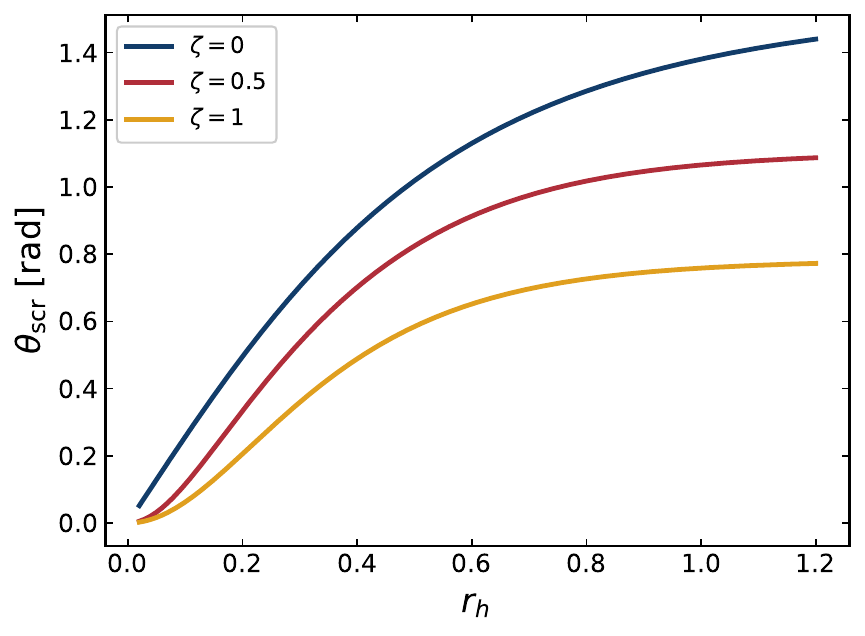}\hfill
\includegraphics[width=0.48\textwidth]{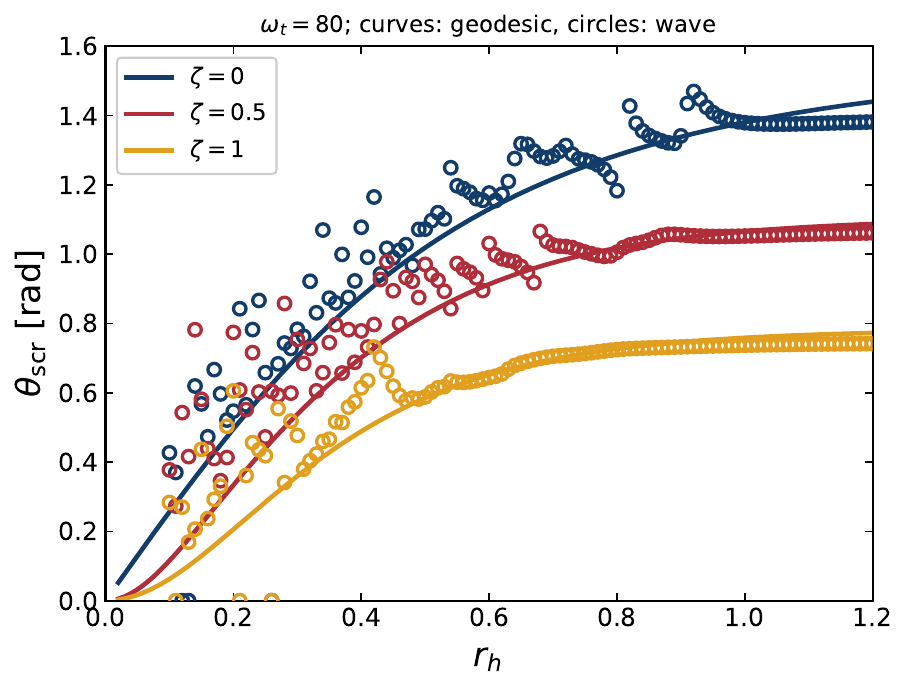}
\caption{Left: geodesic coordinate screen angle
$\theta_{\rm scr}=\arcsin b_t$ from Eq.\eqref{eq:ring} for
$\zeta=0,0.5,1$. The curves approach
$\arcsin[(1+\zeta^2)^{-1/2}]$, whereas the incidence
angle in the normalized boundary frame in Eq.\eqref{eq:boundaryangle} approaches $\pi/2$ for every $\zeta$.
Right: comparison of the wave optics coordinate angle with the geodesic
curves. Circles denote the global brightness maximum for $\omega=80$,
$d=0.5$, and a horizon radius step of $0.01$. The wave data follow the
geodesic result for $r_h\gtrsim0.45$. At smaller $r_h$, the diffraction
fluctuations present already at $\zeta=0$ \cite{Hashimoto:2020mgx} increase
as the ring approaches the aperture scale
$\arcsin[\pi/(\omega d)]\simeq0.079$. Points on the horizontal axis occur
when the central spot is brighter than the ring.}
\label{fig:geo_multi}
\label{fig:compare_multi}
\end{figure*}

\begin{table}[tbp]
\caption{\rev{Representative values of the wave optics coordinate peak position}
$x_{\rm peak}/f_{\rm L}$ against the geodesic prediction
$b_t$ of Eq.\eqref{eq:ring} ($\omega=80$, $d=0.5$,
$\sigma=0.01$).}
\label{tab:checks}
\begin{ruledtabular}
\begin{tabular}{ccccc}
$r_h$ & $\zeta$ & wave & geodesic & $\Delta$ \\
\hline
$0.9$ & $0$   & $0.9738$ & $0.9732$ & $+0.0006$ \\
$0.9$ & $0.5$ & $0.8703$ & $0.8656$ & $+0.0047$ \\
$0.9$ & $1$   & $0.6686$ & $0.6788$ & $-0.0102$ \\
$0.6$ & $0$   & $0.9232$ & $0.9044$ & $+0.0188$ \\
$0.6$ & $0.5$ & \rev{$0.8576$} & $0.7915$ & $+0.0661$ \\
$0.6$ & $1$   & $0.6003$ & $0.6067$ & $-0.0064$ \\
$0.3$ & $0$   & $0.7056$ & $0.6475$ & $+0.0581$ \\
$0.3$ & $0.5$ & $0.6846$ & $0.5124$ & $+0.1722$ \\
$0.3$ & $1$   & $0.4588$ & $0.3519$ & \rev{$+0.1069$} \\
\end{tabular}
\end{ruledtabular}
\end{table}

To quantify the ring radius, we scan $r_h\in[0.1,1.2]$ in steps of
$0.01$ for each $\zeta$ and construct the image at
$\theta_{\rm obs}=0$. The global brightness maximum along the $x_S$ axis is
located with parabolic subpixel refinement, and the coordinate screen angle
is $\theta_{\rm scr}=\arcsin(x_{\rm peak}/f_{\rm L})$. The geodesic curves
\eqref{eq:ring} are shown in Fig.~\ref{fig:geo_multi}, the wave optics
comparison in Fig.~\ref{fig:compare_multi}, and representative values in
Table~\ref{tab:checks}.
\rev{At $(r_h,\zeta)=(0.81,0)$, the maximum occurs at
$x_{\rm peak}/f_{\rm L}=1.0068$, outside the unit angular chart.
This value is retained in the numerical data but is not assigned a
real angle in Fig.~\ref{fig:compare_multi}.}

\rev{The scan illustrates the effects of finite frequency, aperture
resolution, and boundary normalization.}

(a) Quantitative agreement in the coordinate screen.
For $r_h\gtrsim0.45$ \redrev{the wave optics points follow the geodesic curves for
the corresponding values of} $\zeta$. Typical residuals are a few percent, with the
largest relative deviations \rev{in $x_{\rm peak}/f_{\rm L}$ from $b_t$,
over $r_h\geq0.45$,} reaching about $13\%$; they
oscillate around zero in the same manner as the fluctuation induced by diffraction observed in
the classical background at $\omega=80$ \cite{Hashimoto:2020mgx}. At large
$r_h$ the three families flatten onto distinct coordinate plateaus: for
$\zeta=1$ the measured value is
$\theta_{\rm scr}\simeq0.74$--$0.78$, against
$\arcsin(1/\sqrt{2})=0.785$.  This agreement tests the use of $\omega_t$ in
the lens kernel.  It should not be interpreted as a physical incidence
angle until the factor $\sqrt{A}$ in Eq.\eqref{eq:boundaryangle} is applied.

(b) Observable window at small $r_h$.
The optical system resolves angular scales down to
$\Delta(x_S/f_{\rm L})\simeq\pi/(\omega d)$
[Eq.\eqref{eq:resolution}], equal to $0.079$ for our fiducial parameters.
The coordinate shrinking pushes the ring of small black holes toward, and
eventually below, this scale [for example, $b_t=0.062$ at
$(r_h,\zeta)=(0.1,1)$], where the peak position decouples from the geodesic
prediction: the requirement
\begin{equation}
b_t\;\gtrsim\;\frac{\pi}{\omega d}
\quad\Longleftrightarrow\quad
w_{\max}\big(1+\zeta^{2}w_{\max}\big)\;\lesssim\;
\Big(\frac{\omega d}{\pi}\Big)^{2}
\label{eq:window}
\end{equation}
\rev{gives an approximate resolution criterion for the ring at fixed
$(\omega,d)$. The criterion alone does not guarantee that the ring is the
global brightness maximum.} In addition, at isolated small horizon radii
(present already at $\zeta=0$, e.g.\ $r_h\simeq0.12$--$0.13$, and at
$r_h\simeq0.11,\,0.21,\,0.26$ for $\zeta=1$) the interference pattern makes
\rev{a central intensity maximum} \redrev{brighter than the ring}, so the global
maximum jumps to the screen center \rev{as shown by points on the
horizontal axis in Fig.~\ref{fig:compare_multi}. Increasing $\omega d$
improves aperture resolution, but the relative brightness of competing
peaks remains sensitive to the frequency and the response.}

(c) Boundary normalization and thermal comparison.
The coordinate temperature $T_t(r_h)$ in Eq.\eqref{eq:temperature} carries
no $\zeta$ dependence, but the normalized boundary temperature
$T_{\rm bdy}=T_t/\sqrt{A}$ does.  It is therefore not correct to describe
different values of $\zeta$ at fixed $r_h$ as isothermal states on the same
boundary metric.  In the normalized frame the ring ratio is
Eq.\eqref{eq:boundaryratio}, not Eq.\eqref{eq:ratio}.  Between
$\zeta=0$ and $1$ \rev{the normalized scale $\sin\vartheta_c$} decreases by about $5.1\%$ at
$r_h=0.6$ and by only $0.9\%$ at $r_h=1$, and the difference vanishes as
$r_h\to\infty$.  The larger $30$--$33\%$ shift visible in the raw screen
coordinate remains a valid result for the specific lens convention
\eqref{eq:lens}, but it combines the finite radius photon potential change
with the asymptotic time normalization.

\section{\rev{Photon sphere scales and the WKB approximation}}
\label{sec:wkb}

\rev{The maximum of the null potential determines the coordinate ring
scale in Sec.~\ref{sec:geodesics}. We now calculate the corresponding
height and curvature of the scalar wave barrier using the WKB expansion
\cite{Schutz:1985km,Iyer:1986np,Konoplya:2003ii,Matyjasek:2017psb,
Konoplya:2019hlu}. These quantities yield the orbital frequency and the
Lyapunov exponent of the circular null orbit. Their relation to the
quasinormal spectrum also depends on the reflecting AdS boundary, which
we incorporate through an approximate spectral model.}

\subsection{Barrier data and \rev{their} dependence on $\zeta$ in closed form}

The radial equation \eqref{eq:radial}, rewritten for
$\Phi=e^{-i\Omega t}Y_{lm}\,\psi(r)/r$ in the tortoise coordinate
$\mathrm{d}r_*=\mathrm{d}r/f$, takes the Schr\"odinger form
$\big[\partial_{r_*}^{2}+\Omega^{2}-V(r)\big]\psi=0$ with
\begin{equation}
V(r)=f(r)\left[\frac{l(l+1)}{r^{2}}+\frac{f'(r)}{r}\right]
=l(l+1)\,v(r)+\frac{f f'}{r},
\label{eq:Vwkb}
\end{equation}
\rev{where $v=f/r^{2}$ is the geodesic potential \eqref{eq:veff} and
primes in this section denote $r$ derivatives. The AdS contribution to
$V$ can prevent the formation of a local maximum at small angular momentum
\cite{Festuccia:2008zx,Hashimoto:2023buz}. Let $l_{\min}$ denote the
smallest integer multipole for which the photon sphere barrier exists;
the relevant range is then $l\geq l_{\min}$. For $r_h=0.6$,
$l_{\min}=8,\,7,\,8,\,11$ at $\zeta=0,\,0.5,\,1,\,2$, respectively.
For $r_h=1$, the corresponding values are $45,\,45,\,51,\,74$.
Once the barrier is present, its peak approaches $r_{\rm ps}$ as
$l\to\infty$.}

\rev{We denote the peak position by $r_0$, its height by $V_0=V(r_0)$,
and its tortoise derivatives by
$V_k=(\mathrm{d}^kV/\mathrm{d}r_*^k)_{r_0}$. Higher order WKB
approximations use these local derivatives
\cite{Schutz:1985km,Iyer:1986np,Konoplya:2003ii,Konoplya:2019hlu}.
The usual barrier quantization imposes waves propagating away from the
barrier on both sides. In AdS, however, the outer boundary reflects the
wave. We therefore distinguish the local barrier data from the global
quasinormal spectrum. The model in Sec.~\ref{sec:adsspec} retains the
quadratic expansion specified by $V_0$ and $V_2$ and includes a Dirichlet
wall following Ref.~\cite{Hashimoto:2023buz}.}

\begin{figure*}[t]
\centering
\includegraphics[width=0.96\textwidth]{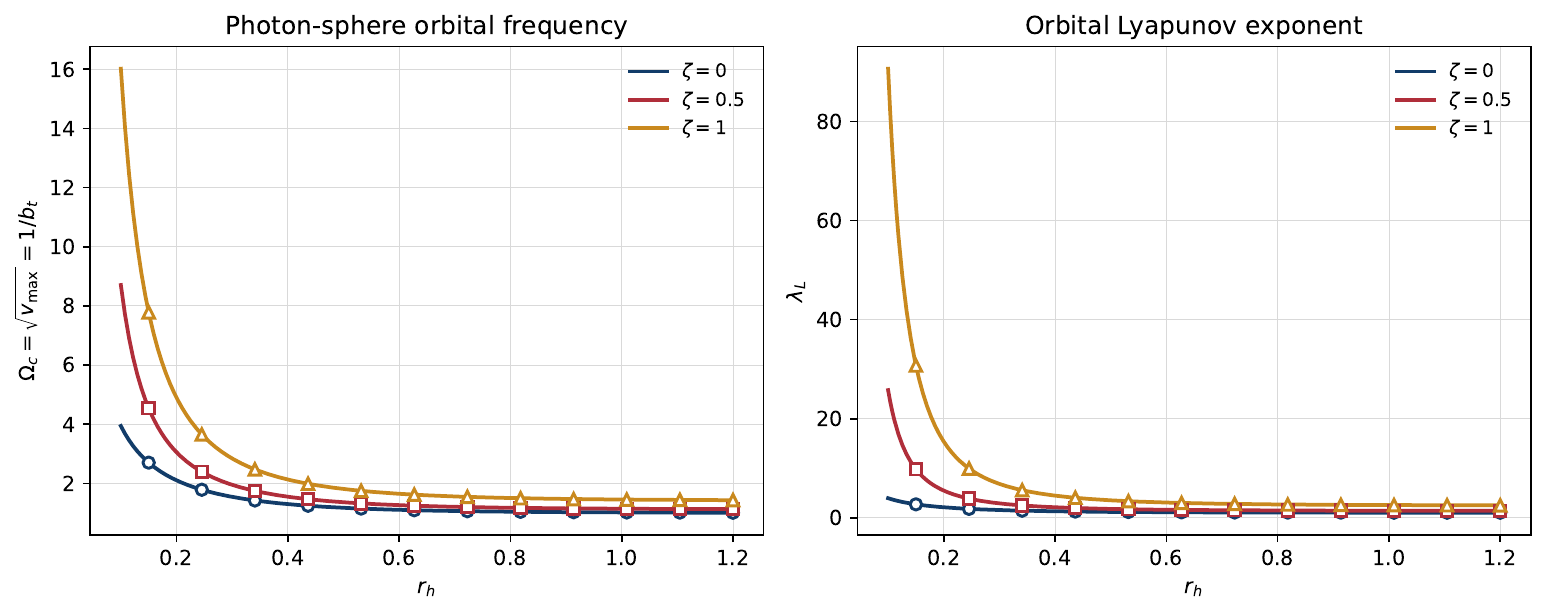}
\caption{Orbital frequency $\Omega_c=1/b_t$ in the left panel and
Lyapunov exponent $\lambda_L$ in the right panel as functions of the horizon
radius for $\zeta=0,\,0.5,\,1$. Curves show the closed forms
\eqref{eq:lyapunov}, and open symbols show the eikonal WKB barrier data at
$l=4\times10^{3}$. The two panels coincide at $\zeta=0$ according to
Eq.\eqref{eq:degeneracy}; for $\zeta>0$, $\lambda_L$ increases faster than
$\Omega_c$ as described by Eq.\eqref{eq:breaking}.}
\label{fig:lyapunov}
\end{figure*}

\begin{figure}[t]
\centering
\includegraphics[width=0.46\textwidth]{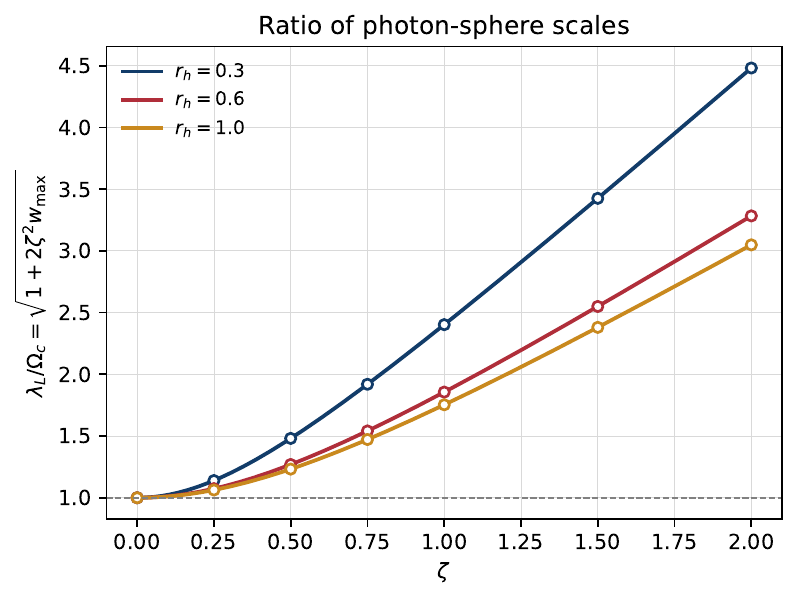}
\caption{\rev{Ratio $\lambda_L/\Omega_c$ as a function of $\zeta$ for
three horizon radii. Curves show Eq.\eqref{eq:breaking}, and symbols
show numerical barrier data. The dashed line indicates the
Schwarzschild--AdS$_4$ value. The ratio determines $\zeta^2w_{\max}$;
extracting $\zeta$ from it requires the horizon radius.}}
\label{fig:ratio}
\end{figure}

\rev{The relation $v=w+\zeta^{2}w^{2}$ and the extremum condition
$w'(r_{\rm ps})=0$ give}
\begin{align}
v''\big|_{\rm ps}&=\big(1+2\zeta^{2}w_{\max}\big)\,w''\big|_{\rm ps},\nonumber\\
f\big|_{\rm ps}&=\rev{F_0(r_{\rm ps})}\big(1+\zeta^{2}w_{\max}\big),
\label{eq:chain2}
\end{align}
\rev{where $F_0(r_{\rm ps})=9M^{2}w_{\max}$. The orbital frequency
and the Lyapunov exponent of the unstable circular null orbit, both
defined with respect to $t$, are \cite{Cardoso:2008bp}}
\begin{equation}
\Omega_c=\sqrt{v_{\max}},\qquad
\lambda_L=\sqrt{-\frac{f^{2}v''}{2v}}\,\bigg|_{\rm ps}\rev{.}
\end{equation}
\rev{Substitution yields}
\begin{equation}
\begin{aligned}
\Omega_c&=\sqrt{w_{\max}}\sqrt{1+\zeta^{2}w_{\max}},\\
\lambda_L&=\sqrt{w_{\max}}\sqrt{\big(1+\zeta^{2}w_{\max}\big)
\big(1+2\zeta^{2}w_{\max}\big)}\,.
\end{aligned}
\label{eq:lyapunov}
\end{equation}
\rev{Thus $\Omega_c=1/b_t$, whereas
$\Omega_{c,{\rm bdy}}=\Omega_c/\sqrt{A}=1/\sin\vartheta_c$ and
$\lambda_{L,{\rm bdy}}=\lambda_L/\sqrt{A}$. The familiar estimate
$\Omega\simeq(l+\tfrac12)\Omega_c-i(n+\tfrac12)\lambda_L$
describes the local barrier model with outgoing waves; its imaginary part is
not a quantization rule for the Dirichlet AdS problem. At $\zeta=0$,
$\lambda_L^{2}=w_{\max}$ and hence}
\begin{equation}
\lambda_L=\Omega_c \qquad (\zeta=0),
\label{eq:degeneracy}
\end{equation}
\rev{for every horizon radius. The quantum correction changes this
ratio to}
\begin{equation}
\frac{\lambda_L}{\Omega_c}=\sqrt{1+2\zeta^{2}w_{\max}}\;,
\label{eq:breaking}
\end{equation}
\rev{which exceeds unity when $\zeta\neq0$. Figures~\ref{fig:lyapunov}
and \ref{fig:ratio} compare these expressions with barrier data evaluated
at $l=4\times10^{3}$. The sampled values reproduce the dependence on
$r_h$ and $\zeta$ predicted by Eqs.\eqref{eq:lyapunov} and
\eqref{eq:breaking}.}

\rev{At fixed $r_h$, both $b_t(\zeta)/b_t(0)$ and
$\lambda_L/\Omega_c$ depend on the product $\zeta^2w_{\max}$.
Eliminating this product gives}
\begin{equation}
\left(\frac{\lambda_L}{\Omega_c}\right)^{2}
=\frac{2}{b_t^{2}\,w_{\max}}-1
=2\,\frac{b_t^{2}(0)}{b_t^{2}(\zeta)}-1,
\label{eq:consistency}
\end{equation}
\rev{This is a geometric identity for the effective metric considered
here. The ratio $\lambda_L/\Omega_c$ is invariant under the rescaling
to boundary time. Its extraction from an AdS wave spectrum requires a
model that also accounts for boundary reflection.}

\subsection{\rev{Approximate spectrum with a reflecting boundary}}
\label{sec:adsspec}

\rev{Following Ref.~\cite{Hashimoto:2023buz}, we approximate the
potential near $r_0$ by an inverted harmonic oscillator and represent the
AdS boundary by a Dirichlet wall. Define
$x=(-2V_2)^{1/4}[r_*-r_*(r_0)]$ and
$E=(\Omega^2-V_0)/\sqrt{-2V_2}$. Near the barrier top,
$\Omega\simeq a+bE$, where $a=\sqrt{V_0}$ and
$b=\sqrt{-V_2/(2V_0)}$; the latter tends to $\lambda_L$ at large $l$.
The dimensionless wall position is
$x_L=(-2V_2)^{1/4}\int_{r_0}^{\infty}\mathrm{d}r/f(r)$.
For $x_L\gg1$, the model gives the quantization condition}
\begin{equation}
\frac{\pi}{2}+\frac{x_L^{2}}{2}
+\Big(2\ln x_L+\frac{i\pi}{2}\Big)E
+i\ln\frac{\sqrt{2\pi}}{\Gamma(\tfrac12-iE)}\in2\pi\mathbf{Z}.
\label{eq:quantization}
\end{equation}
\rev{The parameters are evaluated from the full effective metric, but
the quadratic potential and hard wall remain approximations to the
global radial problem. For $r_h=0.6$ and $l=30$, the tortoise distances
from $r_0$ to the boundary are $0.7352$, $0.5763$, and $0.3495$ at
$\zeta=0$, $0.5$, and $1$, respectively. The asymptotic contribution to
this distance scales as $A^{-1}$, since $f(r)\sim Ar^2$.}

\begin{figure*}[t]
\centering
\includegraphics[width=0.96\textwidth]{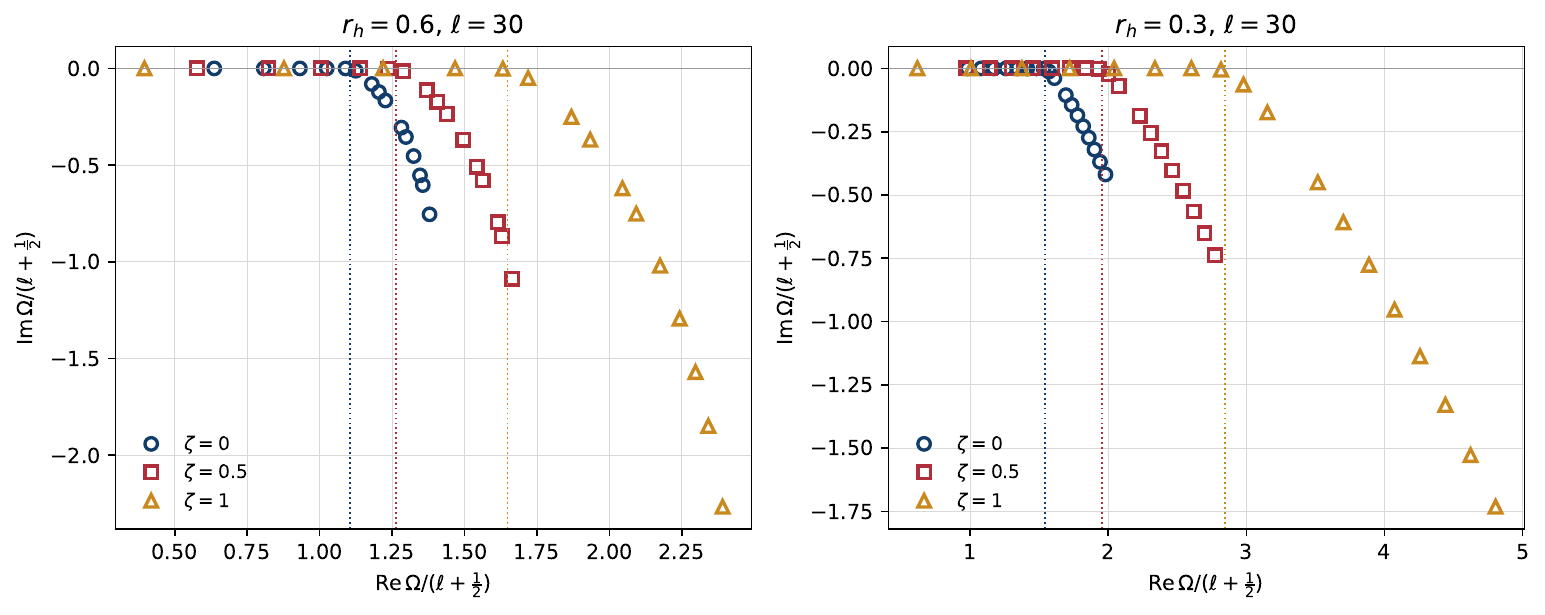}
\caption{\rev{Approximate spectra from the inverted oscillator model
with a Dirichlet wall, Eq.\eqref{eq:quantization}, at $\ell=30$ for
$\zeta=0,\,0.5,\,1$. Frequencies are divided by $\ell+\tfrac12$.
The left and right panels show $r_h=0.6$ and $r_h=0.3$, respectively.
Dotted lines mark $\Omega_c=1/b_t$. The model exhibits weakly damped
states below the barrier and increasing damping near and above it.
Both the barrier curvature and the wall position affect this spectrum.}}
\label{fig:spectra}
\end{figure*}

\rev{Figure~\ref{fig:spectra} shows the resulting spectra at $l=30$.
The states below the barrier are long lived because their leakage toward
the horizon is suppressed, as in Schwarzschild--AdS
\cite{Festuccia:2008zx,Hashimoto:2023buz}. Their damping is small but
nonzero. The transition to stronger damping occurs near
$\mathrm{Re}\,\Omega\simeq(l+\tfrac12)\Omega_c$, linking the spectral
transition to the coordinate ring scale. The complex spectral locus
depends on $x_L$ as well as on $\lambda_L$, so its slope cannot be
identified with the Lyapunov exponent alone.}

\begin{table}[b]
\caption{\rev{Comparison of the finite multipole barrier estimate
$b_t^{\rm WKB}=(l+\tfrac12)/\sqrt{V_0}$ at $l=30$ with the
geodesic value and the wave optics peak from Table~\ref{tab:checks}.
The horizon radius is $r_h=0.6$, and the image frequency is $\omega=80$.
The last column gives the geometric ratio in Eq.\eqref{eq:breaking}.}}
\label{tab:wkbring}
\begin{ruledtabular}
\begin{tabular}{ccccc}
$\zeta$ & WKB $(l{=}30)$ & geodesic & wave optics & $\lambda_L/\Omega_c$ \\
\hline
$0$   & $0.9028$ & $0.9044$ & $0.9232$ & $1.000$ \\
$0.5$ & $0.7896$ & $0.7915$ & \rev{$0.8576$} & $1.269$ \\
$1$   & $0.6041$ & $0.6067$ & $0.6003$ & $1.856$ \\
\end{tabular}
\end{ruledtabular}
\end{table}

\rev{In Table~\ref{tab:wkbring}, the barrier estimates differ from the
geodesic values by $0.18\%$--$0.42\%$. With $q=l+\tfrac12$, the potential
height has the expansion $V_0=q^2v_{\max}+\mathcal{O}(1)$, so
$q/\sqrt{V_0}=b_t[1+\mathcal{O}(q^{-2})]$. The wave optics offsets are
larger and include the finite frequency and aperture effects discussed
in Sec.~\ref{sec:radius}. The barrier height therefore supplies a local
geometric reference for both the image and the approximate spectrum.}

\section{\rev{Time domain modes and the spectral transition}}
\label{sec:prony}

\rev{We also evolve the scalar master equation on an ingoing null grid
and extract damped exponentials using least squares Prony and
matrix pencil methods \cite{Konoplya:2011qq}. Candidate frequencies are
compared across shifted fitting windows and several model orders to
identify roots sensitive to the fit. In the asymptotically flat
Schwarzschild benchmark, the fundamental frequencies for
$\ell=0,\ldots,3$ agree with published values to within $0.4\%$.
For the AdS background, an initial pulse modulated by a carrier resolves
19 trapped modes at $(r_h,\zeta,\ell)=(0.3,0,56)$. Their median relative
difference from the frequency domain poles is $0.069\%$; grid refinement
is consistent with second order convergence.}

\rev{The change in damping near the barrier top provides a spectral
estimate of the ring scale. We fit the locus of the frequency domain
poles to the hard wall model of Ref.~\cite{Hashimoto:2023buz}, allowing
the barrier height, curvature scale, and wall position to vary. The
fitted heights at $\ell=40$ and $56$ estimate the leading coefficient
of $\ell(\ell+1)$ through}
\begin{equation}
v_{\rm knee}=\frac{V_{\rm fit}(56)-V_{\rm fit}(40)}
{56\times57-40\times41},
\qquad
\sin\vartheta_c^{\rm knee}=\sqrt{\frac{A}{v_{\rm knee}}}.
\label{eq:pronyknee}
\end{equation}
\rev{For $r_h=0.3$, the sampled values are}
\begin{center}
\begin{ruledtabular}
\begin{tabular}{cccc}
$\zeta$ & \rev{spectral knee} & geodesic & deviation \\
\hline
$0$   & $0.647418(72)$ & $0.647459$ & $-0.006\%$ \\
$0.5$ & $0.572909(35)$ & $0.572928$ & $-0.003\%$ \\
$1$   & $0.497810(11)$ & $0.497642$ & $+0.034\%$ \\
\end{tabular}
\end{ruledtabular}
\end{center}
\rev{The entries are estimates of $\sin\vartheta_c$ in the normalized
boundary frame. Parentheses give the numerical uncertainty in the last
digits, combining the fit covariance and the change between two fitting
windows. They do not include the full systematic error of the spectral
model. The deviations from Eq.\eqref{eq:boundaryangle} range from
$-0.006\%$ to $+0.034\%$ at the three sampled values of $\zeta$.}

\begin{figure*}[tbp]
\centering
\includegraphics[width=0.48\textwidth]{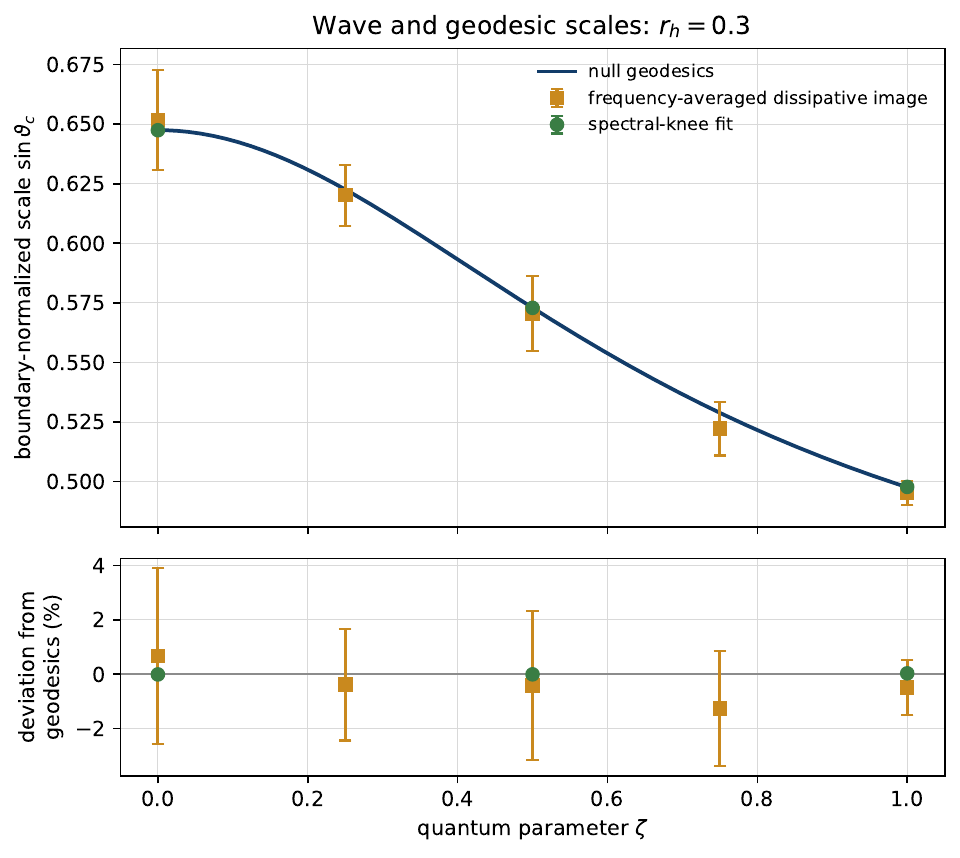}
\caption{\rev{Ring estimates in the normalized boundary frame for $r_h=0.3$.
The spectral knee points follow the geodesic prediction, with the
numerical fit uncertainties defined in the text. The image points use
the dissipative response averaged over
$\omega_{\rm bdy}=76,78,80,82,84$; their error bars show the variation
among the peak positions at individual frequencies. Their offsets from the
geodesic curve are smaller than this spread with frequency.}}
\label{fig:pronyring}
\end{figure*}

\rev{Figure~\ref{fig:pronyring} also includes images formed from the
dissipative response $i\,\mathrm{Im}\,O_t$, where $O_t$ is the cubic
boundary coefficient after removing the local Eddington--Finkelstein
contact terms. The five boundary frequencies are
$\omega_{\rm bdy}=76,78,80,82,84$. At each frequency the intensity
profile is normalized to its maximum; the peak of the averaged profiles
defines the plotted ring position. The image error bars show the
standard deviation of the peak positions at the five individual frequencies.
This frequency average uses the dissipative response, whereas
Sec.~\ref{sec:images} uses the full complex response at a single
coordinate frequency.}

\rev{The time domain comparison tests the frequency domain pole
calculation. The subsequent knee fit additionally assumes the
spectral model near the barrier. Its agreement with the geodesic scale is
therefore a consistency test within that approximation.}

\section{Optical appearance: backward ray tracing}
\label{sec:optical}

\rev{Backward ray tracing gives a geometric optics image of a specified
emitting disk \cite{Gralla:2019xty,Yang:2022btw}. We use it to distinguish
the critical null orbit, which is fixed by the metric, from the
brightness distribution, which also depends on the source model.
All calculations in this section use a static observer at $r_o=200$.}

\subsection{Orbit number and ring classification}

\rev{For a null ray with coordinate impact parameter $b=L/E_t$,
$(\mathrm{d}u/\mathrm{d}\varphi)^2=b^{-2}-v(1/u)$.
A returning ray has $b_c<b<[v(r_o)]^{-1/2}$ and turns at the largest
root $r_t$ of $v(r_t)=b^{-2}$. Its \redrev{total azimuthal angle} between the inward and
outward crossings of the observer sphere is}
\begin{equation}
\varphi(b)=2\!\int_{1/r_o}^{1/r_t}
\frac{\mathrm{d}u}{\sqrt{b^{-2}-v(1/u)}}.
\label{eq:deflection}
\end{equation}
\rev{The orbit number $n=\varphi/(2\pi)$ counts complete revolutions.
We use $n=3/4$ and $n=5/4$ to define the outer lensed ring and photon ring
windows \cite{Gralla:2019xty}. Their widths characterize the returning
branch. Captured rays with $b<b_c$ are also included when computing disk
emission: they can intersect the emitting region before reaching the
horizon.}

The essential difference from the asymptotically flat case is geometric:
in the qSAdS background $v(r)$ rises from zero at the horizon to its maximum
$v_{\max}$ at the photon sphere. Its radius is independent of $\zeta$
and is given by $r_{\rm ps}=3M$. The potential then decreases toward the AdS boundary, where $v\to1+\zeta^2$. The
critical impact parameter is therefore
\begin{equation}
b_c^{(t)}=\frac{1}{\sqrt{v_{\max}}}
=\big[\,w_{\max}(1+\zeta^2w_{\max})\,\big]^{-1/2}=b_t,
\label{eq:bc}
\end{equation}
identical to the geodesic coordinate scale \eqref{eq:ring} and to the
eikonal $1/\Omega_c$ of Eq.\eqref{eq:lyapunov}. The equality follows
because all three quantities are determined by the maximum of the same
potential. In the flat Schwarzschild limit, the ray tracing calculation
recovers $r_{\rm ps}=3M$ and $b_c=3\sqrt3\,M=5.196\,M$, together with the
standard boundaries $n=5/4$ at $b=5.228\,M$ and $n=3/4$ at
$b=6.168\,M$ \cite{Gralla:2019xty}.

\begin{figure*}[tbp]
\centering
\includegraphics[width=0.96\textwidth]{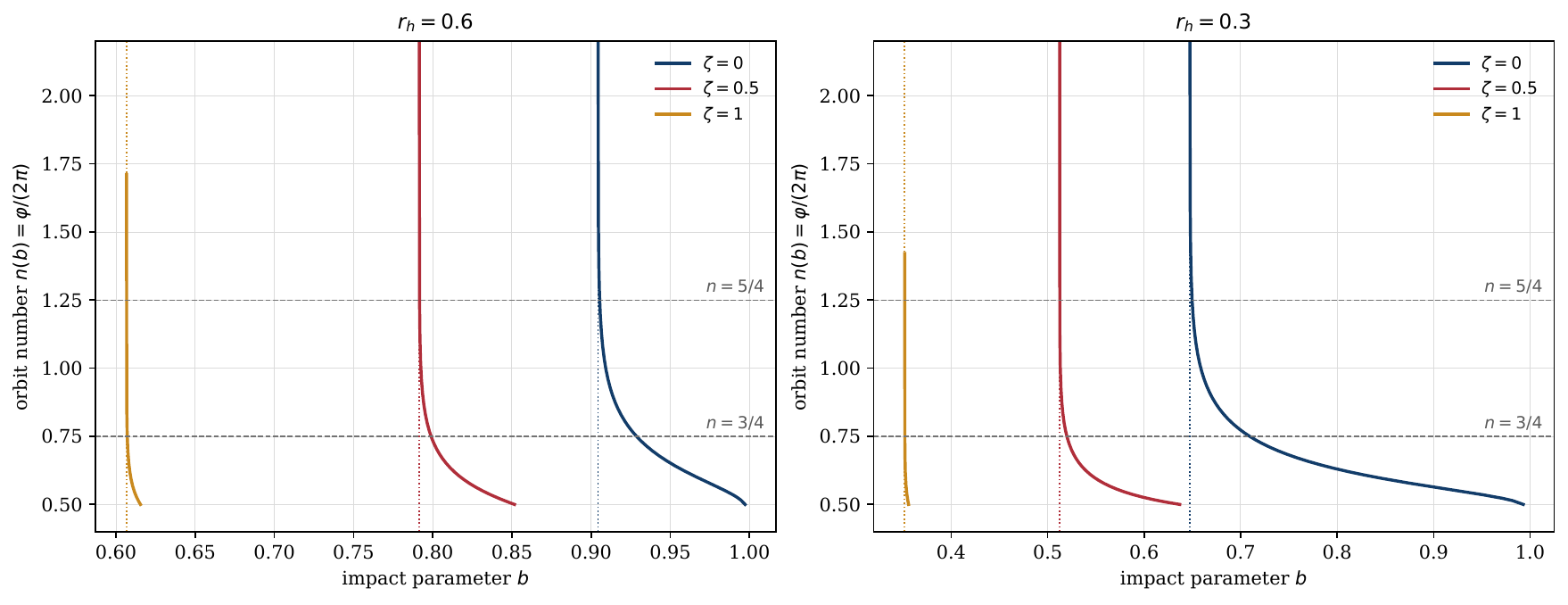}
\caption{Total number of \rev{complete revolutions} $n(b)=\varphi(b)/2\pi$ as a
function of impact parameter for $\zeta=0,\,0.5,\,1$. The left and right
panels correspond to $r_h=0.6$ and $r_h=0.3$. Dotted vertical lines mark
$b_c=b_t$, where the curves diverge as $b\to b_c^{+}$. The dashed lines at
$n=3/4$ and $n=5/4$ separate direct emission, the lensed ring, and the
photon ring. Increasing $\zeta$ shifts the curves to smaller $b$ and narrows
the impact parameter intervals occupied by the lensed and photon rings.
\rev{Only $n\geq1/2$ is displayed; the segment near the observer edge is excluded
from the plot, not from the geodesic calculation.}}
\label{fig:norbits}
\end{figure*}

\subsection{Quantum shrinking of the ring windows}

Fig~\ref{fig:norbits} shows $n(b)$ for two representative horizon
radii. In addition to shifting $b_c$ according to Eq.\eqref{eq:bc}, \redrev{increasing the
quantum parameter narrows} the impact parameter windows of the higher order
images. Table~\ref{tab:optical} lists the photon ring width
$\Delta b_{\rm PR}=b_{5/4}-b_c$ and the lensed ring width
$\Delta b_{\rm LR}=b_{3/4}-b_{5/4}$, where $b_{5/4}$ and $b_{3/4}$ solve
$n=5/4$ and $n=3/4$, respectively. Both widths decrease by more than an
order of magnitude between $\zeta=0$ and $\zeta=1$. At $r_h=0.6$, the
photon ring width decreases from $9.8\times10^{-4}$ to below $10^{-5}$,
while the lensed ring width decreases from $2.3\times10^{-2}$ to
$4.5\times10^{-4}$. The contraction is monotonic and becomes stronger for
smaller black holes, consistent with the $r_h$ dependence found in
Sec.~\ref{sec:radius}.

\begin{table}[b]
\caption{Photon ring and lensed ring coordinate impact parameter windows of
the qSAdS black hole from backward ray tracing: $b_c^{(t)}=b_t$, and the widths
$\Delta b_{\rm PR}=b_{5/4}-b_c$, $\Delta b_{\rm LR}=b_{3/4}-b_{5/4}$.
The quantum parameter both shifts $b_c$ inward and compresses the ring
windows.}
\label{tab:optical}
\begin{ruledtabular}
\begin{tabular}{cccccc}
$r_h$ & $\zeta$ & $b_c$ & $b_{5/4}$ & $\Delta b_{\rm PR}$ & $\Delta b_{\rm LR}$ \\
\hline
$0.6$ & $0$   & $0.9044$ & $0.9054$ & $9.8\times10^{-4}$ & $2.33\times10^{-2}$ \\
$0.6$ & $0.5$ & $0.7915$ & $0.7917$ & $1.3\times10^{-4}$ & $7.30\times10^{-3}$ \\
$0.6$ & $1$   & $0.6067$ & $0.6067$ & \rev{$1.30\times10^{-6}$} & $4.5\times10^{-4}$ \\
$0.3$ & $0$   & $0.6475$ & $0.6497$ & $2.3\times10^{-3}$ & \rev{$5.96\times10^{-2}$} \\
$0.3$ & $0.5$ & $0.5124$ & $0.5125$ & $7\times10^{-5}$   & $7.46\times10^{-3}$ \\
$0.3$ & $1$   & $0.3519$ & $0.3519$ & \rev{$4.64\times10^{-8}$} & \rev{$8.80\times10^{-5}$} \\
\end{tabular}
\end{ruledtabular}
\end{table}

\begin{figure*}[tbp]
\centering
\includegraphics[width=0.96\textwidth]{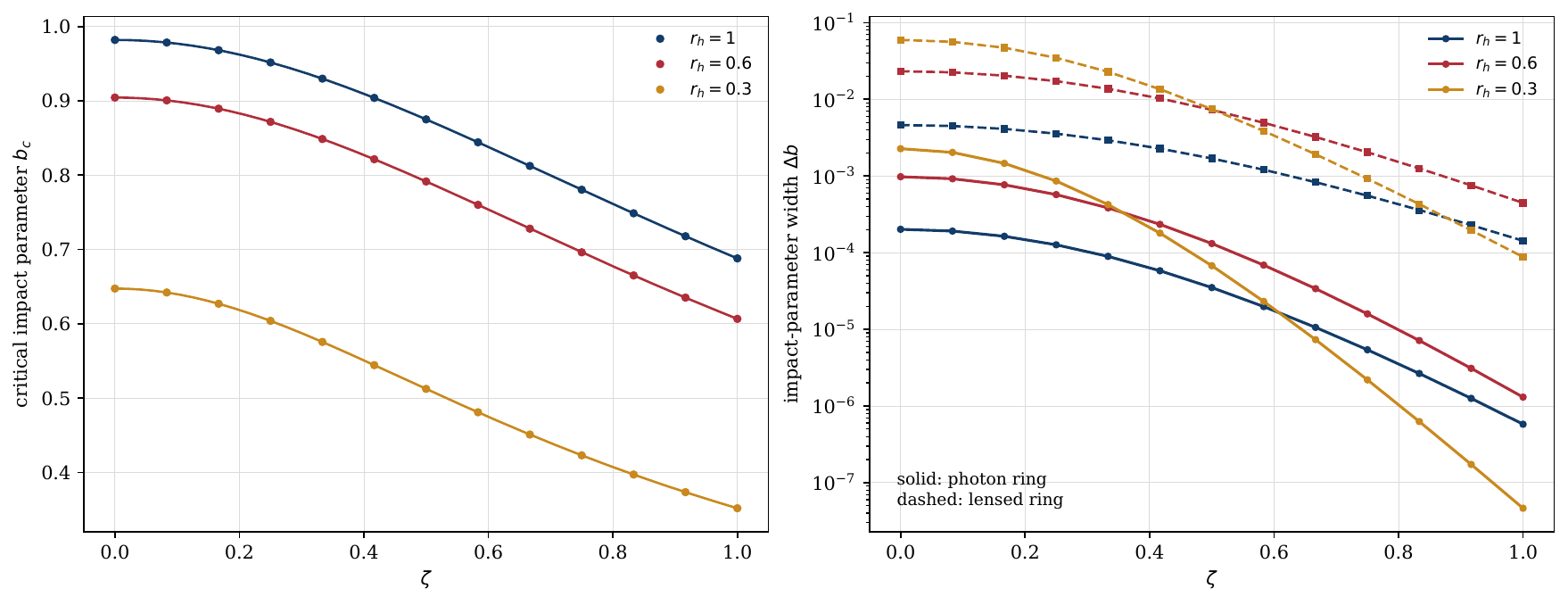}
\caption{Left: critical impact parameter $b_c(\zeta)$ for three horizon
radii. \redrev{Symbols show the ray tracing values, and curves show the closed}
form \eqref{eq:bc}. Right: photon ring width $\Delta b_{\rm PR}$ and
lensed ring width $\Delta b_{\rm LR}$ on a logarithmic scale. Solid and
dashed curves represent the two widths, respectively; both decrease by more
than an order of magnitude as $\zeta\to1$.}
\label{fig:optwindows}
\end{figure*}

\subsection{Observed appearance}

\rev{We assume a static, optically thin, isotropically emitting disk
viewed along its symmetry axis. The invariant $I_\nu/\nu^3$ gives}
\begin{equation}
\rev{I_{\nu,o}(b)=\sum_m g_m^3\,
  I_{\nu,e}\!\left(r_m,\frac{\nu_o}{g_m}\right),\qquad
g_m=\sqrt{\frac{f(r_m)}{f(r_o)}}.}
\label{eq:disktransfer}
\end{equation}
\rev{Here $r_m$ denotes a disk crossing along the backward ray.
We include crossings on both inward and outward segments of returning
rays, as well as crossings of captured rays. For a spectrally flat
emitted specific intensity over the relevant frequency interval, we use}
\begin{equation}
\begin{gathered}
\rev{I_{\nu,e}(r)=C\,r^{-2}\exp[-(r/R_d)^2]\,H(r-r_{\rm ps}),}\\
\rev{R_d=10r_{\rm ps}.}
\end{gathered}
\label{eq:emissivity}
\end{equation}
\rev{The normalization $C$ cancels when each image is normalized.
The smooth outer taper is part of the emission model. It is required
because, in AdS, $g^3/r^2$ grows proportionally to $r$ at large emission
radius; an untapered disk would give a brightness controlled by its
outer cutoff. Eq.\eqref{eq:disktransfer} concerns specific
intensity. An intensity integrated over frequency would instead acquire
the redshift factor $g^4$.}

\begin{figure*}[tbp]
\centering
\includegraphics[width=0.96\textwidth]{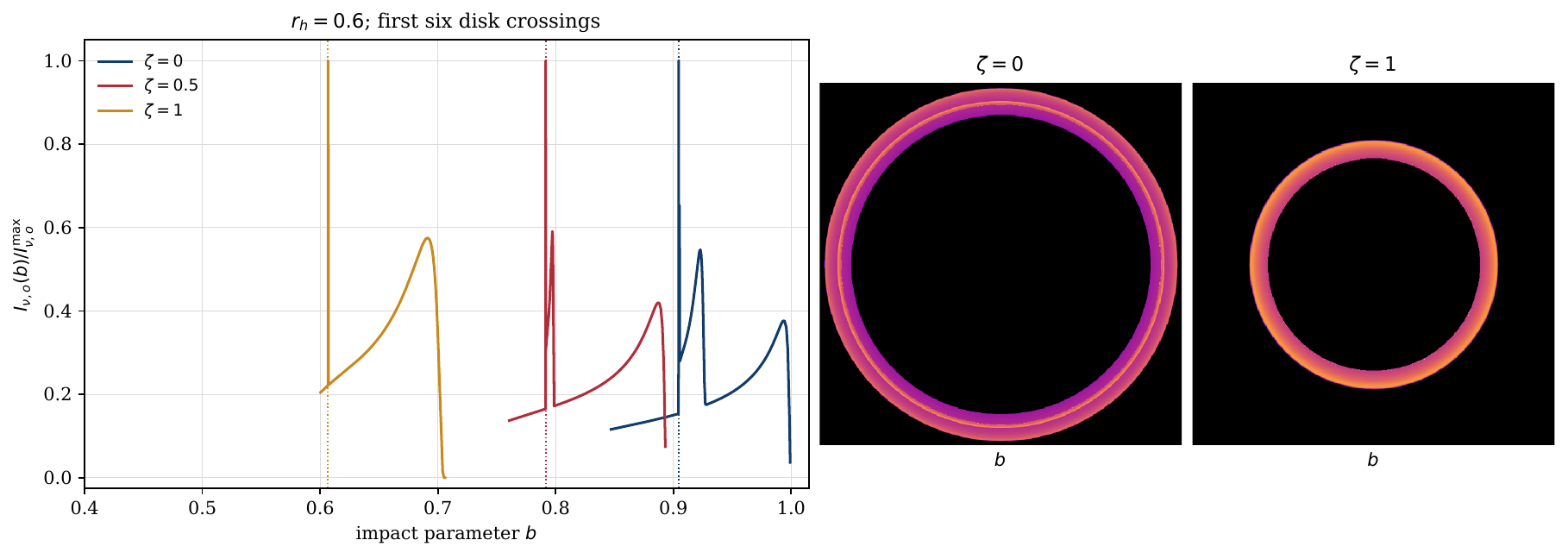}
\caption{\rev{Specific intensity images for $r_h=0.6$ and $r_o=200$,
using the disk model in Eq.\eqref{eq:emissivity}. The left panel shows
individually normalized profiles for $\zeta=0,0.5,1$, \redrev{plotted only where
the emission is nonzero, without connecting lines to the region with no emission}; dotted lines mark $b_c$. The right panels show the $\zeta=0$ and $\zeta=1$ images on a
square root intensity scale. Both captured and returning rays contribute
to the emission. The broad brightness distribution depends on the
emissivity, while the critical curve is determined by the metric.}}
\label{fig:optimage}
\end{figure*}

\rev{Figure~\ref{fig:optimage} shows the resulting profiles and
axisymmetric images. The broad emission depends on the disk model,
whereas the narrow higher order features accumulate near $b_c$.
Their critical positions shift inward as $\zeta$ increases.
A sharp feature associated with the disk's inner edge is physically
distinct from a numerical integration artifact.}

\rev{The null orbit scale, the leading local barrier frequency, and
the critical curve obtained by ray tracing satisfy $b_t=b_c^{(t)}=1/\Omega_c$.
This identity does not require a disk model. By contrast, a brightness
maximum or a global AdS quasinormal frequency need not coincide with
the geometric value. Conversion of coordinate impact parameters to the
normalized boundary frame multiplies them by $\sqrt{A}$.}

\section{Pseudospectral quasinormal modes}
\label{sec:pseudospectral}

\rev{The local WKB barrier analysis and hard wall approximation of
Sec.~\ref{sec:wkb} do not solve the full radial boundary value problem.
We therefore compute scalar quasinormal frequencies by Chebyshev
collocation \cite{Jansen:2017oag}. This calculation imposes regularity at
the future horizon and a vanishing source at the AdS boundary, including
at multipoles for which no photon sphere barrier exists.}

\subsection{Method and validation}

Writing the massless scalar field in ingoing Eddington--Finkelstein form,
$\Phi=e^{-i\omega v}\,S(r)\,Y_{\ell m}$ with $v=t+r_*$, turns the QNM
boundary conditions into \rev{regularity at the future horizon}
and a Dirichlet condition at the AdS boundary. The radial
equation is
\begin{equation}
r^2 f\,S''+\big(r^2 f'+2rf-2i\omega r^2\big)S'
-\big(\ell(\ell+1)+2i\omega r\big)S=0 .
\label{eq:EF}
\end{equation}
\rev{We compactify the exterior with $x=r_h/r$, so that the boundary and
horizon lie at $x=0$ and $x=1$. \redrev{Factoring out the normalizable behavior by writing}
$S=x^3W$ imposes the condition of vanishing source and removes the singular
boundary coefficients. Chebyshev--Gauss--Lobatto collocation gives a
matrix pencil linear in $\omega$,
$(\mathbb{M}_0+\omega\mathbb{M}_1)\mathbf{W}=0$.
Only roots that agree under grid refinement are retained. Roots with large frequencies that move with the grid are excluded before any ordering or
mode selection.}

\rev{Two selection rules serve different comparisons. At low multipole we
report the lowest oscillatory modes, independently of the geometric
frequency. For the coordinate scale comparison in
Sec.~\ref{sec:synthesis}, we select the converged oscillatory root whose
real part is nearest to $(\ell+\tfrac12)\Omega_c$. This second rule uses
the geometric prediction and \redrev{can select different overtones as parameters vary}.
It is not a continuation of one fixed mode branch.}

At $\zeta=0$, the background reduces to Schwarzschild--AdS$_4$, whose
scalar spectrum provides a benchmark. For $r_h=1$ and $\ell=0$, the first
three modes are $\omega_0=2.79822-2.67121\,i$,
$\omega_1=4.75849-5.03757\,i$, and
$\omega_2=6.71927-7.39449\,i$, in agreement with the continued fraction
results of Berti, Cardoso, and Starinets \cite{Berti:2009kk}.

\begin{table}[tbp]
\caption{\rev{Lowest oscillatory scalar mode retained after grid
refinement at $\ell=2$. The geometric frequency $\Omega_c=1/b_t$ is
included for comparison. At this multipole the scalar potential has no
photon sphere barrier for these backgrounds, so the entries are not
identified as photon sphere modes.}}
\label{tab:pspec}
\begin{ruledtabular}
\begin{tabular}{cccc}
$r_h$ & $\zeta$ & \rev{$\omega_{\rm osc}\ (\ell=2)$} & $\Omega_c=1/b_t$ \\
\hline
$1$   & $0$   & $4.1171-2.2756\,i$ & $1.0184$ \\
$1$   & $0.5$ & $5.0924-3.3333\,i$ & $1.1428$ \\
$1$   & $1$   & $8.1686-6.1700\,i$ & $1.4534$ \\
$0.6$ & $0$   & $3.9829-1.0841\,i$ & $1.1057$ \\
$0.6$ & $0.5$ & $4.9874-1.5510\,i$ & $1.2634$ \\
$0.6$ & $1$   & $7.9021-2.8845\,i$ & $1.6483$ \\
\end{tabular}
\end{ruledtabular}
\end{table}

\subsection{\rev{Spectrum and comparison with the photon sphere scale}}

\begin{figure*}[tbp]
\centering
\includegraphics[width=0.9\textwidth]{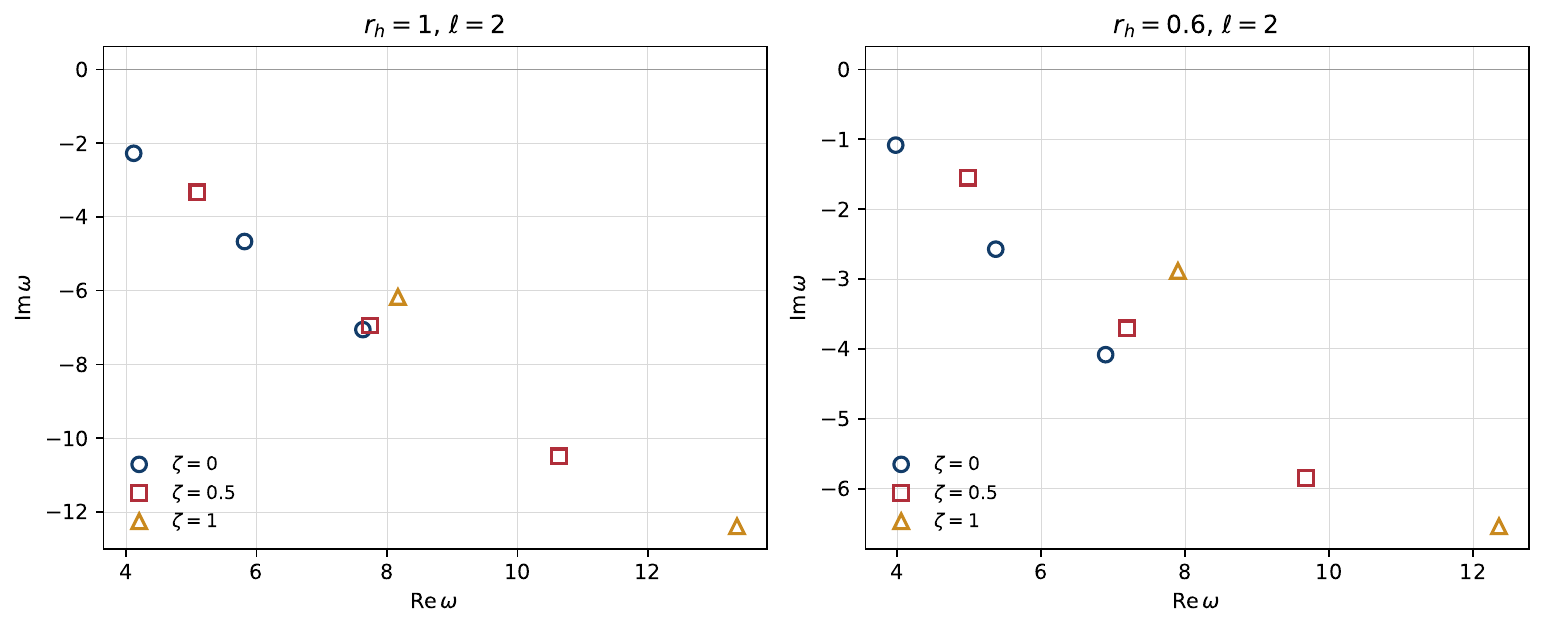}
\caption{\rev{Lowest oscillatory scalar quasinormal
frequencies that converge under grid refinement at $\ell=2$ for $\zeta=0,0.5,1$. The left and right panels
show $r_h=1$ and $r_h=0.6$. The displayed roots are filtered by agreement
between collocation grids before they are ordered by frequency.}}
\label{fig:pspecspectrum}
\end{figure*}

\rev{Table~\ref{tab:pspec} and Fig.~\ref{fig:pspecspectrum} show the
lower part of the oscillatory spectrum. \redrev{The real parts of the frequencies and the damping rates} of the
listed modes increase with $\zeta$. This trend is obtained from the full
AdS boundary value problem and does not require a local barrier.
The orbital Lyapunov exponent alone does not determine these damping
rates.}

\begin{figure}[tbp]
\centering
\includegraphics[width=0.46\textwidth]{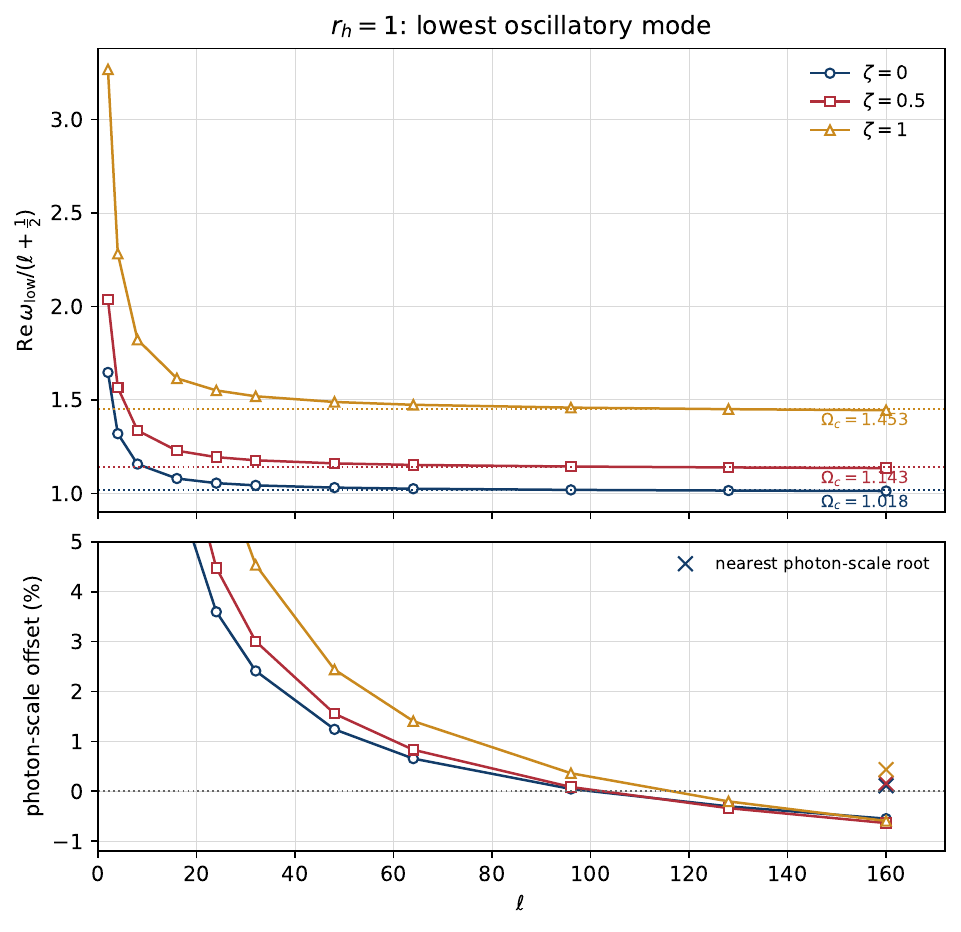}
\caption{\rev{Multipole dependence of oscillatory scalar frequencies at
$r_h=1$, compared with the geometric values $\Omega_c=1/b_t$.
The extended range distinguishes the lowest oscillatory branch from
roots selected by proximity to the photon sphere frequency. The
horizontal lines are geometric reference scales, not assumed limits
of every mode branch.}}
\label{fig:pspeceikonal}
\end{figure}

\rev{Figure~\ref{fig:pspeceikonal} compares the ratio
$\mathrm{Re}\,\omega_{\rm osc}/(\ell+\tfrac12)$ with $\Omega_c$ at
$r_h=1$. A decrease toward the geometric line over a finite multipole
range does not establish an asymptotic limit. In particular, for
$\zeta=0$ the lowest oscillatory branch crosses below $\Omega_c$ when the
range is extended to higher $\ell$. A root chosen at each $\ell$ by
proximity to $\Omega_c$ can instead switch to a different overtone.
These are distinct spectral constructions. The reflecting AdS boundary
supports trapped modes \cite{Festuccia:2008zx}; it prevents the
asymptotically flat photon sphere correspondence from being applied to
every AdS mode \cite{Cardoso:2008bp}.}

\rev{The exact identity $b_t=b_c^{(t)}=1/\Omega_c$ relates geometric
quantities and the leading local barrier scale. Quasinormal frequencies
also depend on the boundary condition and mode index. The pseudospectral
calculation tests finite multipole spectral behavior without replacing
these dependencies by a geometric identity. All coordinate frequencies
are converted to the normalized boundary frame by division by
$\sqrt{A}$.}

\section{\rev{Comparison of geometric and spectral scales}}
\label{sec:synthesis}

\rev{We compare the geometric impact parameter $b_t=1/\Omega_c$ with
the spectral estimate $(\ell+\tfrac12)/\mathrm{Re}\,\omega$ in a common
coordinate convention. Geodesics, leading eikonal WKB barrier data, and
ray tracing use the same potential maximum. Their equality is therefore
a geometric identity. The wave spectrum supplies additional information
about the AdS boundary condition and mode dependence. Its comparison
with $b_t$ requires an explicit selection rule, as described in
Sec.~\ref{sec:pseudospectral}.}

\subsection{\rev{Geometric identity and finite multipole offsets}}

Table~\ref{tab:master} \redrev{compares four determinations of} $b_t$ \redrev{on a common parameter grid}.
The knee analysis based on the Prony method is shown separately in the normalized boundary
frame in Fig.~\ref{fig:pronyring}. The geodesic, WKB, and optical values
coincide because each reduces to
$b_t=[\,w_{\max}(1+\zeta^2w_{\max})]^{-1/2}$ in the relevant limit. Thus the
WKB barrier top, the ray tracing critical curve, and the geodesic photon
sphere \redrev{yield the same critical impact parameter}.

\begin{table}[b]
\caption{\rev{Coordinate geometric scale and finite multipole spectral
estimates. G denotes the analytic null orbit value, W the leading eikonal
barrier value, and O the ray tracing critical value; these share the same
potential maximum. P is $(\ell+\tfrac12)/\mathrm{Re}\,\omega$ at
$\ell=16$ for the converged root nearest to
$(\ell+\tfrac12)\Omega_c$ in real frequency. The P column therefore uses
a selection rule informed by the geometry.}}
\label{tab:master}
\begin{ruledtabular}
\begin{tabular}{cccccc}
$r_h$ & $\zeta$ & $b_t^{\rm G}$ & $b_t^{\rm W}$ & $b_t^{\rm O}$ & $b_t^{\rm P}$ \\
\hline
$0.6$ & $0$   & $0.90443$ & $0.90443$ & $0.90443$ & $0.90035$ \\
$0.6$ & $0.5$ & $0.79153$ & $0.79153$ & $0.79153$ & $0.78641$ \\
$0.6$ & $1$   & $0.60668$ & $0.60668$ & $0.60668$ & $0.59313$ \\
$1.0$ & $0$   & $0.98198$ & $0.98198$ & $0.98198$ & $0.92625$ \\
$1.0$ & $0.5$ & $0.87508$ & $0.87508$ & $0.87508$ & $0.81406$ \\
$1.0$ & $1$   & $0.68802$ & $0.68802$ & $0.68802$ & $0.61913$ \\
\end{tabular}
\end{ruledtabular}
\end{table}

\begin{figure*}[t]
\centering
\includegraphics[width=0.92\textwidth]{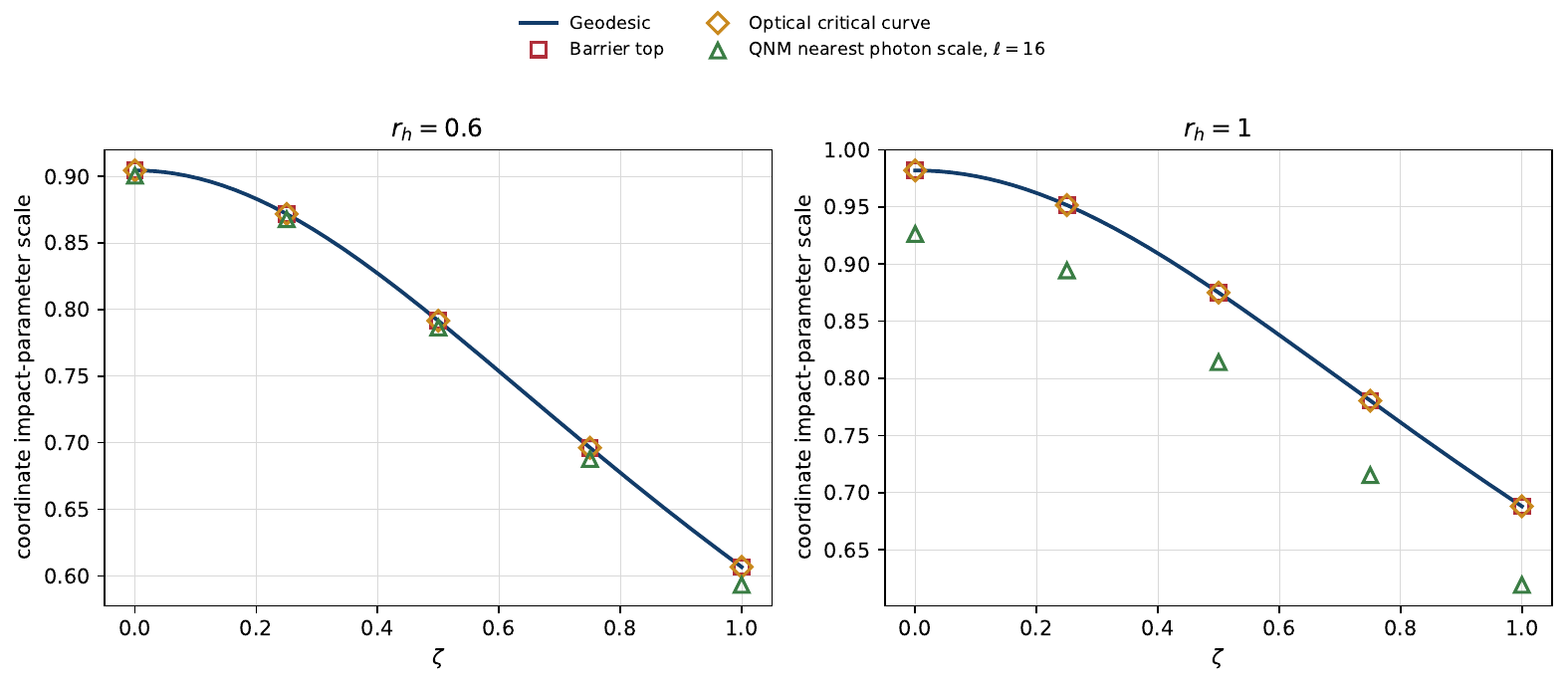}
\caption{\rev{Coordinate scale comparison for $r_h=0.6$ on the left and
$r_h=1$ on the right. The geodesic curve, leading eikonal WKB squares,
and ray tracing diamonds share the same geometric limit. Triangles show
the $\ell=16$ spectral estimate selected by proximity to the geometric
frequency. Their offsets include finite multipole, boundary condition,
and mode selection effects.}}
\label{fig:masterbc}
\end{figure*}

\rev{At $\ell=16$, the scalar potential has a barrier for $r_h=0.6$
at the displayed values of $\zeta$, but not for $r_h=1$.
The spectral offsets in Fig.~\ref{fig:masterbc} must be interpreted with
this difference in mind. Barrier existence alone does not identify a
global AdS mode with the unstable null orbit. In addition, selecting the
root nearest a geometric target limits the independence of the
comparison.}

\begin{figure*}[t]
\centering
\includegraphics[width=0.92\textwidth]{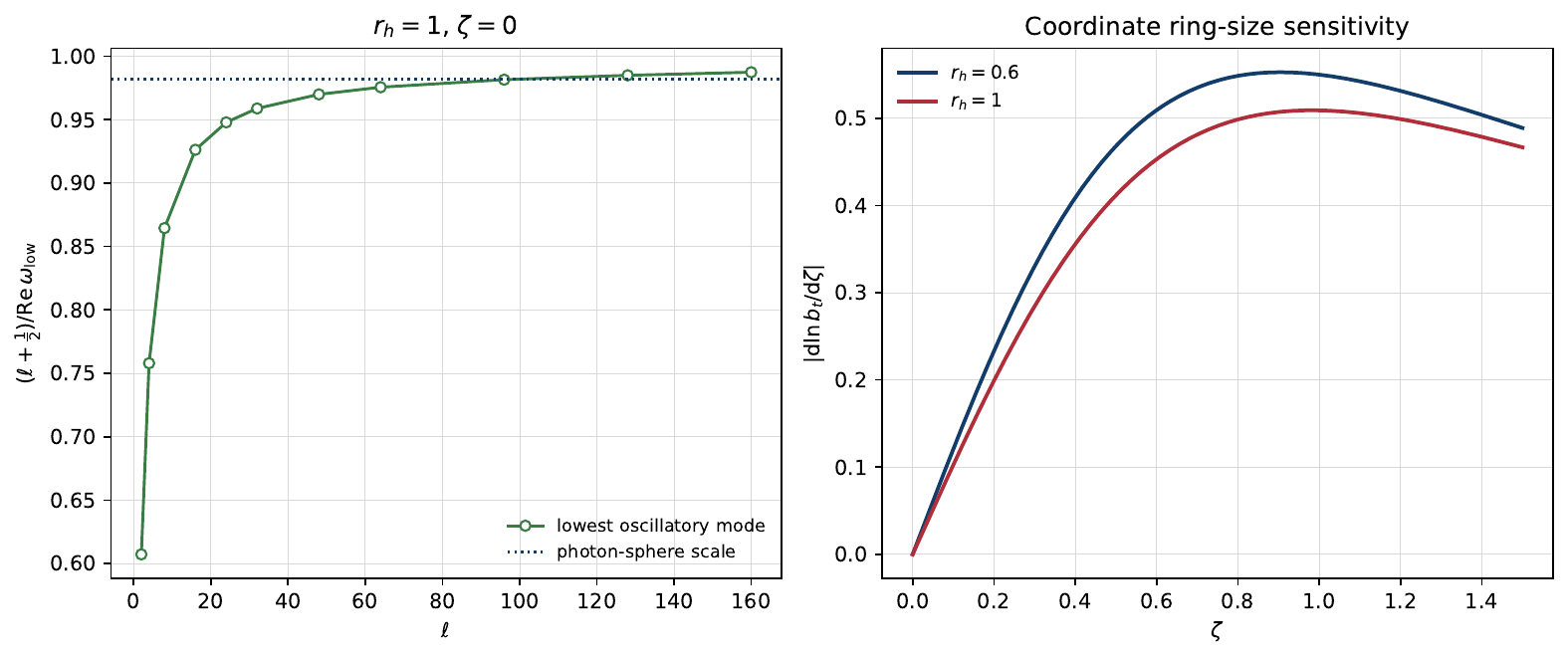}
\caption{\rev{Left: multipole dependence of the scale inferred from
the lowest oscillatory mode at $r_h=1,\zeta=0$, compared with the
geometric value. Extending the range reveals that the two scales cross.
The local barrier first exists at $\ell=45$.} Right: ring size
sensitivity
$|d\ln b_t/d\zeta|=\zeta w_{\max}/(1+\zeta^2 w_{\max})$. The sensitivity
vanishes at $\zeta=0$, reaches its maximum at
$\zeta=1/\sqrt{w_{\max}}$, and then decreases as $1/\zeta$.}
\label{fig:masterconv}
\end{figure*}

\rev{The extended multipole sequence in Fig.~\ref{fig:masterconv}
illustrates this limitation. At $r_h=1,\zeta=0$, the lowest oscillatory
mode initially gives a scale closer to $b_t$ as $\ell$ increases, but
the scale subsequently crosses the geometric value. Thus at low $\ell$ the
\redrev{trend does not establish a photon sphere limit}.
No universal $1/\ell$ correction to the global AdS spectrum is inferred
from this sequence.}

\begin{figure*}[t]
\centering
\includegraphics[width=0.92\textwidth]{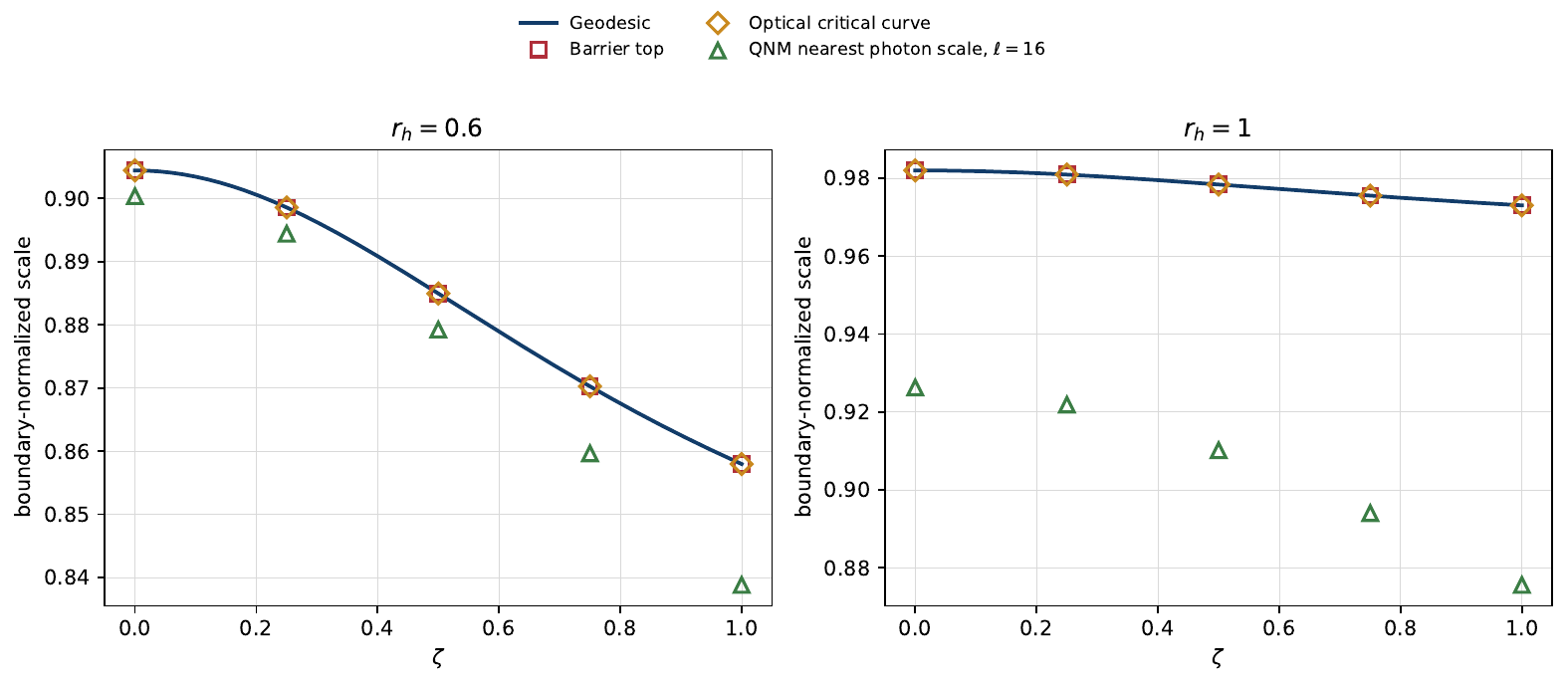}
\caption{The comparison of Fig.~\ref{fig:masterbc} after transforming to the
boundary frame with normalized time, $\widehat b_c=\sqrt{1+\zeta^2}\,b_t=
\sin\vartheta_c$.  The analytic, WKB, and optical values now change only
weakly, especially for $r_h=1$.  The pseudospectral points at finite $\ell$
retain their offset; for $r_h=1$, $\ell=16$ lies below the photon barrier
threshold and should not be interpreted as an eikonal measurement.}
\label{fig:masterboundary}
\end{figure*}

\subsection{\rev{Parameter sensitivity and boundary normalization}}

\rev{The geometric coordinate scale decreases monotonically with
$\zeta$.} Between $\zeta=0$ and $\zeta=1$ the analytic
decrease is $32.9\%$ for $r_h=0.6$ and $29.9\%$ for $r_h=1$. Its logarithmic
sensitivity is
\begin{equation}
\frac{\mathrm{d}\ln b_t}{\mathrm{d}\zeta}=-\frac{\zeta\,w_{\max}}{1+\zeta^2 w_{\max}},
\label{eq:sensitivity}
\end{equation}
\rev{Its magnitude is plotted in the right panel of
Fig.~\ref{fig:masterconv}. It vanishes at $\zeta=0$, reaches its maximum
$\sqrt{w_{\max}}/2$ at $\zeta=1/\sqrt{w_{\max}}$, and then decreases.} For the incidence
angle in the normalized boundary frame, $\sin\vartheta_c=\sqrt{1+\zeta^2}\,b_t$, the corresponding result is
\begin{equation}
\frac{\mathrm{d}\ln(\sin\vartheta_c)}{\mathrm{d}\zeta}
=-\frac{\zeta(w_{\max}-1)}{(1+\zeta^2)(1+\zeta^2w_{\max})}.
\label{eq:boundarysensitivity}
\end{equation}
\rev{The fractional decrease of $\sin\vartheta_c$, shown in
Fig.~\ref{fig:masterboundary}, is $5.1\%$ at $r_h=0.6$ and $0.9\%$ at
$r_h=1$. These percentages refer to the sine of the incidence angle.
The corresponding decreases of $\vartheta_c$ itself are $8.7\%$ and
$3.1\%$.}

\rev{The common $w_{\max}$ dependence explains the agreement among
the geometric determinations. The spectral offsets contain additional
information but are not, by themselves, measurements of a universal
eikonal correction. All comparisons here hold $r_h$ fixed. The coordinate
temperature is then unchanged, while the temperature in the normalized
boundary frame varies according to Eq.\eqref{eq:Tbdy}.
Comparing isothermal states would require a different parameter scan.
The quasinormal spectra in Ref.~\cite{Liu:2025qnm} were computed
in de Sitter spacetime, with outgoing conditions at the cosmological
horizon rather than Dirichlet conditions at the AdS boundary.
No equivalence between individual modes of the two spectra is assumed.}

\section{Conclusions and discussion}
\label{sec:conclusion}

\rev{We have constructed holographic Einstein ring images for a
covariant quantum corrected Schwarzschild--AdS$_4$ geometry.
At fixed horizon radius, the mass parameter $M$, coordinate surface
gravity, and photon sphere radius retain their classical values.
The deformation changes the photon potential height and the asymptotic
lapse coefficient. The resulting critical coordinate scale $b_t$
approaches $(1+\zeta^2)^{-1/2}$ at large $r_h$, whereas the boundary
incidence angle approaches $\pi/2$. Wave optics images show an inward
shift of the ring, with finite frequency offsets. The aperture criterion
in Eq.\eqref{eq:window} estimates whether this scale is resolved; a
brighter central peak can still displace the global intensity maximum.}

\rev{The local WKB barrier data and ray tracing critical curve recover
the same geometric scale. The thin disk brightness additionally depends
on the emissivity model and the observer radius. Our ray tracing model
includes both captured and returning rays, with an explicitly specified
outer emissivity taper. Thus the critical curve is a geometric
quantity, while the broad emission profile is model dependent.}

\rev{The time domain and frequency domain calculations recover
consistent trapped mode poles. Fits to the spectral model near the barrier
give estimates of $\sin\vartheta_c$ within $0.034\%$ of the geodesic
values at $\zeta=0,0.5,1$ and $r_h=0.3$. \redrev{These estimates depend on the assumptions
underlying the spectral model}. The pseudospectral results show that
proximity to the photon sphere frequency over a finite multipole range
does not establish an eikonal limit along a fixed branch. The mode index,
reflecting boundary, and existence of a local barrier must be specified
when interpreting an AdS frequency in terms of a null orbit.}

\rev{Coordinate quantities and those normalized in the boundary frame must be transformed
consistently. A prescribed coordinate frequency and the coordinate
temperature acquire the factor $A^{-1/2}$ on conversion to boundary
time, while the critical angular scale acquires $\sqrt{A}$.
A comparison of deformation parameters must therefore specify the
metric frame, thermal variables, and imaging convention.
The present construction probes an effective bulk geometry;
identifying its microscopic boundary theory remains an open problem.}

\rev{Extensions to the charged solutions
\cite{Yang:2025ufs,Dong:2026qnm} can test how electric charge and the
quantum parameter jointly affect the image. A decomposition of the
boundary Green function into its poles would also clarify which
spectral branches contribute to the ring beyond geometric optics.
Both questions require the full boundary value problem and consistent
normalization of its sources and responses.}

\begin{acknowledgments}
This study is supported in part by the National Natural Science Foundation of
China (Grant No. 12333008).
\end{acknowledgments}

\appendix
\section{Numerical details and convergence tests}
\label{app:numerics}

\rev{Discretization. We discretize Eq.\eqref{eq:radial} on the
Chebyshev--Gauss--Lobatto grid
$u_j=\tfrac{u_h}{2}(\cos\frac{\pi j}{N}+1)$, $j=0,\ldots,N$.
The diagonal entries of each differentiation matrix are chosen so that
its row sums vanish \cite{Boyd:2000,Jansen:2017oag}. The horizon row
imposes the regularity condition obtained from $g(u_h)=0$, and the
boundary row imposes $U_l(0)=1$. We extract $a_3$ by a least squares
projection of
$U_l-[1-i\omega u/g_0-l(l+1)u^2/(2g_0)]$ onto $u^3$ over the eight
interior points nearest the boundary. The source terms use
$g_0=1+\zeta^2$. Eqs.\eqref{eq:a1a2}--\eqref{eq:bexp}
show that the coefficient $a_3$ is not fixed by the local source
expansion and must be obtained from the radial solution.}

\rev{Multipole solution. After eliminating the Dirichlet row,
the angular index enters through the diagonal shift
$[\mathcal M-l(l+1)\operatorname{diag}(u^2)]z=b$.
One eigendecomposition of $\operatorname{diag}(u^{-2})\mathcal M$
therefore supplies the solutions for all multipoles.
Selected modes are also solved by direct LU factorization.
\redrev{Direct solutions are used when the differences exceed
the prescribed tolerance}. This comparison tests the linear solver;
it does not by itself establish convergence of the boundary fit or
the image.}

\rev{Resolution. The imaging calculations use
$N=\lceil\pi\omega/r_h\rceil$, with an upper limit of $1400$ for
the horizon radius scan. We retain multipoles through $l_{\max}=860$
for $\sigma=0.01$. The lens and screen are sampled on $768^2$ and
$1024^2$ grids, respectively, with $|x_S|/f_{\rm L}\leq1.2$.
Peak positions are refined by a local parabolic fit. Increasing $N$
by $20\%$ shifts the peak by $3.1\times10^{-7}$ and
$4.6\times10^{-8}$ at $(r_h,\zeta)=(0.6,0)$ and $(0.3,1)$,
respectively, for $\omega=80$.}

\rev{Independent checks. Numerical extremization of $v(r)$
is compared with Eqs.\eqref{eq:rps} and \eqref{eq:ring}.
The WKB calculation is checked against the analytic large multipole
barrier data. Pseudospectral roots are matched between resolutions
before they enter the figures or tables. The Prony routines are
tested with a synthetic damped signal of known frequencies and with
the Schwarzschild reference spectrum. These checks address different
sources of numerical error and are kept separate from the physical
approximations used to interpret the images and modes.}



\begin{thebibliography}{99}

\bibitem{Maldacena:1997re}
J.~M.~Maldacena,
{The large $N$ limit of superconformal field theories and supergravity},
\href{https://doi.org/10.4310/ATMP.1998.v2.n2.a1}{Adv. Theor. Math. Phys. {\bfseries 2}, no.2, 231--252 (1998)};
\href{https://doi.org/10.1023/A:1026654312961}{Int. J. Theor. Phys. {\bfseries 38}, no.4, 1113--1133 (1999)},
[\href{https://arxiv.org/abs/hep-th/9711200}{arXiv:hep-th/9711200}].

\bibitem{Gubser:1998bc}
S.~S.~Gubser, I.~R.~Klebanov and A.~M.~Polyakov,
{Gauge theory correlators from non-critical string theory},
\href{https://doi.org/10.1016/S0370-2693(98)00377-3}{Phys. Lett. B {\bfseries 428}, no.1--2, 105--114 (1998)},
[\href{https://arxiv.org/abs/hep-th/9802109}{arXiv:hep-th/9802109}].

\bibitem{Witten:1998qj}
E.~Witten,
{Anti de Sitter space and holography},
\href{https://doi.org/10.4310/ATMP.1998.v2.n2.a2}{Adv. Theor. Math. Phys. {\bfseries 2}, no.2, 253--291 (1998)},
[\href{https://arxiv.org/abs/hep-th/9802150}{arXiv:hep-th/9802150}].

\bibitem{Heemskerk:2009pn}
I.~Heemskerk, J.~Penedones, J.~Polchinski and J.~Sully,
{Holography from conformal field theory},
\href{https://doi.org/10.1088/1126-6708/2009/10/079}{JHEP {\bfseries 10}, 079 (2009)},
[\href{https://arxiv.org/abs/0907.0151}{arXiv:0907.0151 [hep-th]}].

\bibitem{EHT:2019dse}
K.~Akiyama \textit{et al.} [Event Horizon Telescope],
{First M87 Event Horizon Telescope Results. I. The Shadow of the Supermassive Black Hole},
\href{https://doi.org/10.3847/2041-8213/ab0ec7}{Astrophys. J. Lett. {\bfseries 875}, no.1, L1 (2019)},
[\href{https://arxiv.org/abs/1906.11238}{arXiv:1906.11238 [astro-ph.GA]}].

\bibitem{Falcke:1999pj}
H.~Falcke, F.~Melia and E.~Agol,
{Viewing the Shadow of the Black Hole at the Galactic Center},
\href{https://doi.org/10.1086/312423}{Astrophys. J. Lett. {\bfseries 528}, no.1, L13--L16 (2000)},
[\href{https://arxiv.org/abs/astro-ph/9912263}{arXiv:astro-ph/9912263}].

\bibitem{Hashimoto:2019jmw}
K.~Hashimoto, S.~Kinoshita and K.~Murata,
{Einstein Rings in Holography},
\href{https://doi.org/10.1103/PhysRevLett.123.031602}{Phys. Rev. Lett. {\bfseries 123}, no.3, 031602 (2019)},
[\href{https://arxiv.org/abs/1906.09113}{arXiv:1906.09113 [hep-th]}].

\bibitem{Hashimoto:2020mgx}
K.~Hashimoto, S.~Kinoshita and K.~Murata,
Imaging black holes through the AdS/CFT correspondence,
\href{https://doi.org/10.1103/PhysRevD.101.066018}{Phys. Rev. D {\bfseries 101}, no.6, 066018 (2020)},
[\href{https://arxiv.org/abs/1811.12617}{arXiv:1811.12617 [hep-th]}].

\bibitem{Nambu:2012wa}
Y.~Nambu,
{Wave optics and image formation in gravitational lensing},
\href{https://doi.org/10.1088/1742-6596/410/1/012036}{J. Phys. Conf. Ser. {\bfseries 410}, 012036 (2013)},
[\href{https://arxiv.org/abs/1207.6846}{arXiv:1207.6846 [gr-qc]}].

\bibitem{Kanai:2013rga}
K.~Kanai and Y.~Nambu,
{Viewing black holes by waves},
\href{https://doi.org/10.1088/0264-9381/30/17/175002}{Class. Quantum Grav. {\bfseries 30}, no.17, 175002 (2013)},
[\href{https://arxiv.org/abs/1303.5520}{arXiv:1303.5520 [gr-qc]}].

\bibitem{Nambu:2015aea}
Y.~Nambu and S.~Noda,
Wave optics in black hole spacetimes: the Schwarzschild case,
\href{https://doi.org/10.1088/0264-9381/33/7/075011}{Class. Quantum Grav. {\bfseries 33}, no.7, 075011 (2016)},
[\href{https://arxiv.org/abs/1502.05468}{arXiv:1502.05468 [gr-qc]}].

\bibitem{Kaku:2021xqp}
Y.~Kaku, K.~Murata and J.~Tsujimura,
Observing black holes through superconductors,
\href{https://doi.org/10.1007/JHEP09(2021)138}{JHEP {\bfseries 09}, 138 (2021)},
[\href{https://arxiv.org/abs/2106.00304}{arXiv:2106.00304 [hep-th]}].

\bibitem{Liu:2022cev}
Y.~Liu, Q.~Chen, X.-X.~Zeng, H.~Zhang and W.~Zhang,
{Holographic Einstein ring of a charged AdS black hole},
\href{https://doi.org/10.1007/JHEP10(2022)189}{JHEP {\bfseries 10}, 189 (2022)},
[\href{https://arxiv.org/abs/2201.03161}{arXiv:2201.03161 [hep-th]}].

\bibitem{Zeng:2023fqy}
X.-X.~Zeng, K.-J.~He, J.~Pu, G.-P.~Li and Q.-Q.~Jiang,
{Holographic Einstein rings of a Gauss--Bonnet AdS black hole},
\href{https://doi.org/10.1140/epjc/s10052-023-12079-5}{Eur. Phys. J. C {\bfseries 83}, no.10, 897 (2023)},
[\href{https://arxiv.org/abs/2302.03692}{arXiv:2302.03692 [gr-qc]}].

\bibitem{Wang:2025image}
Y.-Z.~Wang, G.-P.~Li and K.-J.~He,
{Imaging the quantum corrected AdS black hole via the AdS/CFT correspondence},
\href{https://doi.org/10.1142/S0217732324501955}{Mod. Phys. Lett. A {\bfseries 40}, no.1, 2450195 (2025)}.

\bibitem{Zeng:2025lqg}
X.-X.~Zeng, M.~I.~Aslam, R.~Saleem and X.-Y.~Hu,
{Schwarzschild-like AdS black holes: holographic imaging and LQG influences on Einstein rings},
\href{https://doi.org/10.1140/epjc/s10052-025-14366-9}{Eur. Phys. J. C {\bfseries 85}, no.6, 663 (2025)}.

\bibitem{Qi:2025as}
D.-J.~Qi, M.~Zhang, G.-P.~Li and H.-L.~Li,
{Holographic Einstein ring in asymptotically safe gravity},
\href{https://doi.org/10.1140/epjc/s10052-025-14994-1}{Eur. Phys. J. C {\bfseries 85}, no.11, 1263 (2025)}.

\bibitem{Penrose:1964wq}
R.~Penrose,
{Gravitational Collapse and Space-Time Singularities},
\href{https://doi.org/10.1103/PhysRevLett.14.57}{Phys. Rev. Lett. {\bfseries 14}, no.3, 57--59 (1965)}.

\bibitem{Thiemann:2007pyv}
T.~Thiemann,
{Modern Canonical Quantum General Relativity},
\href{https://doi.org/10.1017/CBO9780511755682}{Cambridge University Press, Cambridge (2007)}.

\bibitem{Rovelli:2004tv}
C.~Rovelli,
{Quantum Gravity},
\href{https://doi.org/10.1017/CBO9780511755804}{Cambridge University Press, Cambridge (2004)}.

\bibitem{Ashtekar:2018lag}
A.~Ashtekar, J.~Olmedo and P.~Singh,
{Quantum Transfiguration of Kruskal Black Holes},
\href{https://doi.org/10.1103/PhysRevLett.121.241301}{Phys. Rev. Lett. {\bfseries 121}, no.24, 241301 (2018)},
[\href{https://arxiv.org/abs/1806.00648}{arXiv:1806.00648 [gr-qc]}].

\bibitem{Zhang:2024khj}
C.~Zhang, J.~Lewandowski, Y.~Ma and J.~Yang,
{Black holes and covariance in effective quantum gravity},
\href{https://doi.org/10.1103/PhysRevD.111.L081504}{Phys. Rev. D {\bfseries 111}, no.8, L081504 (2025)},
[\href{https://arxiv.org/abs/2407.10168}{arXiv:2407.10168 [gr-qc]}].

\bibitem{Zhang:2024ney}
C.~Zhang, J.~Lewandowski, Y.~Ma and J.~Yang,
{Black holes and covariance in effective quantum gravity: A solution without Cauchy horizons},
\href{https://doi.org/10.1103/d6ks-d576}{Phys. Rev. D {\bfseries 112}, no.4, 044054 (2025)},
[\href{https://arxiv.org/abs/2412.02487}{arXiv:2412.02487 [gr-qc]}].

\bibitem{Zhang:2025hlm}
C.~Zhang and Z.~Cao,
{Covariant Dynamics from Static Spherically Symmetric Geometries},
\href{https://doi.org/10.1103/6sws-hfj7}{Phys. Rev. Lett. {\bfseries 135}, no.26, 261401 (2025)},
[\href{https://arxiv.org/abs/2506.09540}{arXiv:2506.09540 [gr-qc]}].

\bibitem{Yang:2025ufs}
J.~Yang, C.~Zhang and Y.~Ma,
{Covariant effective spacetimes of spherically symmetric electrovacuum with a cosmological constant},
\href{https://doi.org/10.1103/rvm2-hwx9}{Phys. Rev. D {\bfseries 112}, no.6, 064049 (2025)},
[\href{https://arxiv.org/abs/2503.15157}{arXiv:2503.15157 [gr-qc]}].

\bibitem{Konoplya:2025mvj}
R.~A.~Konoplya and O.~S.~Stashko,
{Probing the effective quantum gravity via quasinormal modes and shadows of black holes},
\href{https://doi.org/10.1103/PhysRevD.111.104055}{Phys. Rev. D {\bfseries 111}, no.10, 104055 (2025)},
[\href{https://arxiv.org/abs/2408.02578}{arXiv:2408.02578 [gr-qc]}].

\bibitem{Liu:2024soc}
W.~Liu, D.~Wu and J.~Wang,
{Light rings and shadows of static black holes in effective quantum gravity},
\href{https://doi.org/10.1016/j.physletb.2024.139052}{Phys. Lett. B {\bfseries 858}, 139052 (2024)},
[\href{https://arxiv.org/abs/2408.05569}{arXiv:2408.05569 [gr-qc]}].

\bibitem{Lin:2025fcz}
J.~Lin, X.~Zhang and M.~Bravo-Gaete,
{Mass inflation and strong cosmic censorship conjecture in the covariant quantum black hole},
\href{https://doi.org/10.1103/n7jv-crs9}{Phys. Rev. D {\bfseries 111}, no.10, 106025 (2025)},
[\href{https://arxiv.org/abs/2412.01448}{arXiv:2412.01448 [gr-qc]}].

\bibitem{Wang:2025ihs}
R.-B.~Wang, L.~You, S.-J.~Ma, J.-B.~Deng and X.-R.~Hu,
{Thermodynamic phase transition and Joule-Thomson expansion of a quantum corrected black hole in AdS spacetime},
\href{https://doi.org/10.1088/1674-1137/ade4a9}{Chin. Phys. C {\bfseries 49}, no.11, 115102 (2025)},
[\href{https://arxiv.org/abs/2504.06907}{arXiv:2504.06907 [gr-qc]}].

\bibitem{Liu:2025qnm}
Y.~Liu and X.~Zhang,
{Quasinormal modes and tidal Love numbers of covariant effective quantum black holes with cosmological constant},
[\href{https://arxiv.org/abs/2509.12013}{arXiv:2509.12013 [gr-qc]}] (2025).

\bibitem{Dong:2026qnm}
Z.~Dong and J.~Yang,
{Quasinormal modes of charged covariant effective black holes with a cosmological constant},
[\href{https://arxiv.org/abs/2604.19577}{arXiv:2604.19577 [gr-qc]}] (2026).

\bibitem{Ban:2025epjc}
Z.~Ban, J.~Chen and J.~Yang,
{Shadows of rotating black holes in effective quantum gravity},
\href{https://doi.org/10.1140/epjc/s10052-025-14614-y}{Eur. Phys. J. C {\bfseries 85}, no.8, 878 (2025)},
[\href{https://arxiv.org/abs/2411.09374}{arXiv:2411.09374 [gr-qc]}].

\bibitem{Berti:2009kk}
E.~Berti, V.~Cardoso and A.~O.~Starinets,
{Quasinormal modes of black holes and black branes},
\href{https://doi.org/10.1088/0264-9381/26/16/163001}{Class. Quantum Grav. {\bfseries 26}, no.16, 163001 (2009)},
[\href{https://arxiv.org/abs/0905.2975}{arXiv:0905.2975 [gr-qc]}].

\bibitem{Gralla:2019xty}
S.~E.~Gralla, D.~E.~Holz and R.~M.~Wald,
{Black hole shadows, photon rings, and lensing rings},
\href{https://doi.org/10.1103/PhysRevD.100.024018}{Phys. Rev. D {\bfseries 100}, no.2, 024018 (2019)},
[\href{https://arxiv.org/abs/1906.00873}{arXiv:1906.00873 [astro-ph.HE]}].

\bibitem{Yang:2022btw}
J.~Yang, C.~Zhang and Y.~Ma,
{Shadow and stability of quantum-corrected black holes},
\href{https://doi.org/10.1140/epjc/s10052-023-11800-8}{Eur. Phys. J. C {\bfseries 83}, no.7, 619 (2023)},
[\href{https://arxiv.org/abs/2211.04263}{arXiv:2211.04263 [gr-qc]}].

\bibitem{Schutz:1985km}
B.~F.~Schutz and C.~M.~Will,
{Black hole normal modes: A semianalytic approach},
\href{https://doi.org/10.1086/184453}{Astrophys. J. Lett. {\bfseries 291}, L33--L36 (1985)}.

\bibitem{Konoplya:2011qq}
R.~A.~Konoplya and A.~Zhidenko,
{Quasinormal modes of black holes: From astrophysics to string theory},
\href{https://doi.org/10.1103/RevModPhys.83.793}{Rev. Mod. Phys. {\bfseries 83}, no.3, 793--836 (2011)},
[\href{https://arxiv.org/abs/1102.4014}{arXiv:1102.4014 [gr-qc]}].

\bibitem{Iyer:1986np}
S.~Iyer and C.~M.~Will,
Black-hole normal modes: A WKB approach. I. Foundations and application of a higher-order WKB analysis of potential-barrier scattering,
\href{https://doi.org/10.1103/PhysRevD.35.3621}{Phys. Rev. D {\bfseries 35}, no.12, 3621--3631 (1987)}.

\bibitem{Konoplya:2003ii}
R.~A.~Konoplya,
Quasinormal behavior of the $D$-dimensional Schwarzschild black hole and the higher order WKB approach,
\href{https://doi.org/10.1103/PhysRevD.68.024018}{Phys. Rev. D {\bfseries 68}, no.2, 024018 (2003)},
[\href{https://arxiv.org/abs/gr-qc/0303052}{arXiv:gr-qc/0303052}].

\bibitem{Matyjasek:2017psb}
J.~Matyjasek and M.~Opala,
{Quasinormal modes of black holes: The improved semianalytic approach},
\href{https://doi.org/10.1103/PhysRevD.96.024011}{Phys. Rev. D {\bfseries 96}, no.2, 024011 (2017)},
[\href{https://arxiv.org/abs/1704.00361}{arXiv:1704.00361 [gr-qc]}].

\bibitem{Konoplya:2019hlu}
R.~A.~Konoplya, A.~Zhidenko and A.~F.~Zinhailo,
{Higher order WKB formula for quasinormal modes and grey-body factors: recipes for quick and accurate calculations},
\href{https://doi.org/10.1088/1361-6382/ab2e25}{Class. Quantum Grav. {\bfseries 36}, no.15, 155002 (2019)},
[\href{https://arxiv.org/abs/1904.10333}{arXiv:1904.10333 [gr-qc]}].

\bibitem{Cardoso:2008bp}
V.~Cardoso, A.~S.~Miranda, E.~Berti, H.~Witek and V.~T.~Zanchin,
{Geodesic stability, Lyapunov exponents, and quasinormal modes},
\href{https://doi.org/10.1103/PhysRevD.79.064016}{Phys. Rev. D {\bfseries 79}, no.6, 064016 (2009)},
[\href{https://arxiv.org/abs/0812.1806}{arXiv:0812.1806 [hep-th]}].

\bibitem{Festuccia:2008zx}
G.~Festuccia and H.~Liu,
{A Bohr-Sommerfeld Quantization Formula for Quasinormal Frequencies of AdS Black Holes},
\href{https://doi.org/10.1166/asl.2009.1029}{Adv. Sci. Lett. {\bfseries 2}, no.2, 221--235 (2009)},
[\href{https://arxiv.org/abs/0811.1033}{arXiv:0811.1033 [gr-qc]}].

\bibitem{Hashimoto:2023buz}
K.~Hashimoto, K.~Sugiura, K.~Sugiyama and T.~Yoda,
{Photon sphere and quasinormal modes in AdS/CFT},
\href{https://doi.org/10.1007/JHEP10(2023)149}{JHEP {\bfseries 10}, 149 (2023)},
[\href{https://arxiv.org/abs/2307.00237}{arXiv:2307.00237 [hep-th]}].

\bibitem{Hawking:1982dh}
S.~W.~Hawking and D.~N.~Page,
{Thermodynamics of black holes in anti-de Sitter space},
\href{https://doi.org/10.1007/BF01208266}{Commun. Math. Phys. {\bfseries 87}, no.4, 577--588 (1983)}.

\bibitem{Ferrari:1984zz}
V.~Ferrari and B.~Mashhoon,
{New approach to the quasinormal modes of a black hole},
\href{https://doi.org/10.1103/PhysRevD.30.295}{Phys. Rev. D {\bfseries 30}, no.2, 295--304 (1984)}.

\bibitem{Hartnoll:2008vx}
S.~A.~Hartnoll, C.~P.~Herzog and G.~T.~Horowitz,
{Building a Holographic Superconductor},
\href{https://doi.org/10.1103/PhysRevLett.101.031601}{Phys. Rev. Lett. {\bfseries 101}, no.3, 031601 (2008)},
[\href{https://arxiv.org/abs/0803.3295}{arXiv:0803.3295 [hep-th]}].

\bibitem{Boyd:2000}
J.~P.~Boyd,
{Chebyshev and Fourier Spectral Methods}, 2nd ed.,
\href{https://store.doverpublications.com/products/9780486411835}{Dover Publications, Mineola, New York (2001)}.

\bibitem{Jansen:2017oag}
A.~Jansen,
{Overdamped modes in Schwarzschild-de Sitter and a Mathematica package for the numerical computation of quasinormal modes},
\href{https://doi.org/10.1140/epjp/i2017-11825-9}{Eur. Phys. J. Plus {\bfseries 132}, no.12, 546 (2017)},
[\href{https://arxiv.org/abs/1709.09178}{arXiv:1709.09178 [gr-qc]}].

\end{thebibliography}
\end{document}